\documentclass[manuscript,acmsmall,natbib=true]{acmart}
\pdfoutput=1

\makeatletter
\renewcommand{\@formatdoi}[1]{\ignorespaces}
\makeatother

\renewcommand\footnotetextcopyrightpermission[1]{}
\acmJournal{PACMPL}
\acmVolume{1}
\acmNumber{CONF} 
\acmArticle{1}
\acmYear{2018}
\acmMonth{1}
\acmDOI{} 
\startPage{1}

\setcopyright{none}
\usepackage{booktabs}   
\usepackage{subcaption} 

\usepackage[utf8]{inputenc}
\usepackage{amsmath}
\usepackage{amssymb}
\usepackage{amsthm}
\usepackage{mathtools}
\usepackage{stmaryrd}
\usepackage{thmtools}
\usepackage{todonotes}
\usepackage{etoolbox}
\usepackage{appendix}

\usepackage{natbib}
\usepackage{hyperref}
\usepackage{cleveref}

\newif\ifdraft\draftfalse

\ifdraft

\newcommand{\lo}[1]{{\color{red}{[{#1}--Luke]}}}

\newcommand{\todoall}[1]{\todo[inline,color=black!30,author=All]{#1}}
\newcommand{\todompj}[1]{\todo[inline,color=yellow!40,author=Michael]{#1}}
\newcommand{\todomario}[1]{\todo[inline,color=blue!40,author=Mario]{#1}}

\else

\newcommand{\lo}[1]{}

\newcommand{\todoall}[1]{}
\newcommand{\todompj}[1]{}
\newcommand{\todomario}[1]{}
\fi
\usepackage{bbm}

\theoremstyle{plain}
\newtheorem{thm}{Theorem}

\theoremstyle{definition}
\newtheorem{prop}[thm]{Proposition}

\theoremstyle{remark}

\theoremstyle{remark}
\newtheorem{corollary}[thm]{Corollary}

\theoremstyle{remark}

\theoremstyle{definition}
\newtheorem{defn}[thm]{Definition}

\newcommand{\defeq}{\coloneqq}
\newcommand{\cat}[1]{\mathbf{#1}}

\newcommand{\powerset}[1]{\mathcal{P}(#1)}
\newcommand{\denote}[1]{\llbracket #1 \rrbracket}
\newcommand{\sem}[1]{\denote{#1}} 
\newcommand{\ev}{\operatorname{ev}}
\newcommand{\id}{\operatorname{id}}
\newcommand{\doubleplus}{\ensuremath{\mathbin{+\mkern-10mu+}}}
\newcommand{\pair}[2]{\left\langle {#1}, {#2} \right\rangle}
\newcommand{\into}[0]{\rightarrow}

\makeatletter
\newcommand\cplussym
{
  \mathpalette\@incircbin
}
\newcommand\@incircbin[2]
{
  \mathbin
  {
    \ooalign{\hidewidth$#1#2$\hidewidth\crcr$#1\bigcirc$}%
  }
}
\makeatother

\newcommand{\cplus}{\oplus}

\newcommand{\cminus}{\ominus}

\newcommand{\splus}{\cdot}
\newcommand{\mzero}{\mathbf{0}}

\newcommand{\cstruct}[3]{(#1,#2,#3)}
\newcommand{\cstr}[1]{\hat{#1}}
\newcommand{\changes}[1]{\Delta #1}
\newcommand{\change}[1]{\delta #1}

\newcommand{\derive}[1]{#1'}
\newcommand{\supderive}[1]{#1_\uparrow}
\newcommand{\supderiveM}[1]{#1_{\uparrow\uparrow}}
\newcommand{\subderive}[1]{#1_\downarrow}
\newcommand{\subderiveM}[1]{#1_{\downarrow\downarrow}}

\newcommand{\NN}{\mathbb{N}}
\newcommand{\ZZ}{\mathbb{Z}}
  
\newcommand{\upsem}[1]{\sem{\updiff{#1}}}
\newcommand{\downsem}[1]{\sem{\downdiff{#1}}}
\newcommand{\bothsem}[1]{\sem{\bothdiff{#1}}}

\newcommand{\semR}[0]{\mathcal{R}}
\newcommand{\diffR}[0]{\change{\semR}}

\newcommand{\kernel}{\sim}
\newcommand{\kernelOrder}{\leq_\kernel}
\newcommand{\pointwiseOrder}{\leq_{pt}}

\newcommand{\ra}{\rightarrow}

\newcommand{\ptfunc}{\Rightarrow_{pt}}

\newcommand{\superpose}{{\bowtie}}

\newcommand{\exponential}[2]{{#1} \Rightarrow {#2}}

\newcommand{\changeOrder}{\leq_\Delta}

\newcommand{\twist}{\cplus_{\bowtie}}
\newcommand{\disjointTimes}{\bowtie}
\newcommand{\updiff}{\Delta}
\newcommand{\downdiff}{\nabla}

\newcommand{\fixpoint}{\mathbf{fix}}
\newcommand{\lfp}{\mathbf{lfp}}
\newcommand{\adjust}{\mathbf{adjust}}
\newcommand{\iter}{\mathbf{iter}}
\newcommand{\nextiter}{\mathbf{recur}}

\newcommand{\Formula}{\mathrm{Formula}}
\newcommand{\Rel}{\cat{Rel}}
\newcommand{\universalRel}{\mathcal{U}}
\newcommand{\consq}{\mathcal{I}}

\newif\ifproofs

\begin{document}

\author{Mario Alvarez Picallo}
\affiliation{
  \position{PhD Student}
  \department{Computer Science}
  \institution{University of Oxford}            
  \country{United Kingdom}                    
}
\email{mario.alvarez-picallo@cs.ox.ac.uk}          

\author{Alex Eyers-Taylor}
\affiliation{
  \position{Research Engineer}
  \institution{Semmle Ltd}            
  \country{United Kingdom}                    
}
\email{alexet@semmle.com}          

\author{Michael Peyton Jones}
\affiliation{
  \position{Research Engineer}
  \institution{Semmle Ltd}            
  \country{United Kingdom}                    
}
\email{michael@semmle.com}          

\author{C.-H.~Luke Ong}
\affiliation{
  \position{Professor of Computer Science}
  \department{Computer Science}
  \institution{University of Oxford}            
  \country{United Kingdom}                    
}
\email{luke.ong@cs.ox.ac.uk}          

\ccsdesc[500]{Theory of computation~Denotational semantics}
\ccsdesc[500]{Software and its engineering~Constraint and logic languages}

\keywords{incremental computation, semantics, dcpos, Datalog}  

\begin{abstract}
 Incremental computation has recently been studied using the concepts of \emph{change
  structures} and \emph{derivatives} of programs, where the derivative of a function allows updating the output
  of the function based on a change to its input.

  We generalise change structures to \emph{change actions},
  and study their algebraic properties. We develop change actions for common structures
  in computer science, including directed-complete partial orders and Boolean algebras.

  We then show how to compute derivatives of fixpoints. This allows us to
  perform incremental evaluation and maintenance of recursively defined
  functions, with particular application to generalised Datalog programs.

  Moreover, unlike previous results, our techniques are \emph{modular} in that
  they are easy to apply both to variants of Datalog and to other programming languages.
\end{abstract}

\title{Fixing incremental computation}
\subtitle{Derivatives of fixpoints, and the recursive semantics of Datalog}

\maketitle
\thispagestyle{empty}

\section{Introduction}
\label{sec:intro}

Consider the following classic Datalog program\footnote{See \citep[][part D]{abiteboul1995foundations} for an introduction to Datalog.},
which computes the transitive closure of an edge relation $e$:
\begin{align*}
  tc(x, y) &\leftarrow e(x, y)\\
  tc(x, y) &\leftarrow e(x, z) \wedge tc(z, y)
\end{align*}

The semantics of Datalog tells us that the denotation of this program is the
least fixpoint of the rule $tc$. Kleene's Theorem tells us that we can 
compute this fixpoint by repeatedly applying the rule until it stops changing, starting from the empty relation. For example, supposing
that $e = \{ (1, 2), (2, 3), (3, 4) \}$, we get the following evaluation trace:
\begin{center}
  \begin{tabular} {p{3.5em} p{10em} p{10em}}
    Iteration & Newly deduced facts & Accumulated data in $tc$ \\
    \toprule
    0 & $\{ \}$ & $\{ \}$\\
    1 & $\{ (1, 2), (2, 3), (3, 4) \}$ & $\{ (1, 2), (2, 3), (3, 4) \}$\\
    2 & $\{ (1, 2), (2, 3), (3, 4),$ $(1, 3), (2, 4) \}$ & $\{ (1, 2), (2, 3), (3, 4),$ $(1, 3), (2, 4) \}$\\
    3 & $\{ (1, 2), (2, 3), (3, 4),$ $(1, 3), (2, 4), (1, 4),(1, 4) \}$ & $\{ (1, 2), (2, 3), (3, 4),$ $(1, 3), (2, 4), (1, 4) \}$\\
    4 & (as above) & (as above) \\
    \bottomrule
  \end{tabular}
\end{center}
\medskip

At this point we have reached a fixpoint, and so we are done.

However, this process is quite wasteful. We deduced the fact $(1,2)$ at every iteration,
even though we had already deduced it in the first iteration. Indeed, for a
chain of $n$ such edges we will deduce $O(n^2)$ facts along the way.

The standard improvement to this evaluation strategy is known as ``semi-naive'' 
evaluation\citep[See][section 13.1]{abiteboul1995foundations}, where we transform 
the program into a \emph{delta} program with two parts:
\begin{itemize}
  \item A \emph{delta} rule that computes the \emph{new} facts at each
    iteration. 
  \item An \emph{accumulator} rule that accumulates the delta at each
    iteration to compute the final result
\end{itemize}
In this case our delta rule is simple: we only get new transitive edges at iteration $n+1$ if we
can deduce them from transitive edges we deduced at iteration $n$.
\begin{align*}
  \Delta tc_{0}(x, y) &\leftarrow e(x, y)\\
  \Delta tc_{i+1}(x, y) &\leftarrow e(x, z) \wedge \Delta tc_i(z, y)\\
  tc_{0}(x, y) &\leftarrow \Delta tc_0(x, y)\\
  tc_{i+1}(x, y) &\leftarrow tc_{i}(x,y) \vee \Delta tc_{i+1}(x,y)
\end{align*}

\begin{center}
  \begin{tabular} {p{3.5em} p{8em} p{10em}}
    Iteration & $\Delta tc_i$ & $tc_i$ \\
    \toprule
    0 & $\{ (1, 2), (2, 3), (3, 4) \}$ & $\{ (1, 2), (2, 3), (3, 4) \}$\\
    1 & $\{ (1, 3), (2, 4) \}$ & $\{ (1, 2), (2, 3), (3, 4),$ $(1, 3), (2, 4) \}$\\
    2 & $\{ (1, 4) \}$ & $\{ (1, 2), (2, 3), (3, 4),$ $(1, 3), (2, 4), (1, 4) \}$\\
    3 & $\{ \}$ & (as above) \\
    \bottomrule
  \end{tabular}
\end{center}
\medskip

This is much better \textemdash{} we have turned a quadratic computation into a
linear one. The delta transformation is a kind of \emph{incremental
  computation}: at each stage we compute the changes in the rule given the previous
changes to its inputs.

But the delta rule translation works only for traditional Datalog. It is common to
liberalise the formula syntax with additional features, such as disjunction,
existential quantification, negation, and aggregation.\footnote{See, for
  example, \citep[LogiQL]{logicbloxWebsite,halpin2014logiql},
  \citep[Datomic]{datomicWebsite},
  \citep[Souffle]{souffleWebsite,scholz2016fast}, and
  \citep[DES]{saenz2011deductive}, which between them have all of these
  features and more. } 
This allows us to write programs like the following, where we compute whether all the
nodes in a subtree given by $child$ have some property $p$:
\begin{align*}
  treeP(x) &\leftarrow p(x) \wedge \neg \exists y . (child(x,y) \wedge \neg treeP(y))
\end{align*}

Here the combination of negation and explicit existential quantification amounts
to recursion through a \emph{universal} quantifier. We would
like to be able to use semi-naive evaluation for this rule too, but the simple delta
transformation is not well defined for the extended program syntax, and it is unclear how to extend it (and the
correctness proof) to handle such cases.\footnote{A simple approach is to just
  keep applying the algorithm of ``replace each recursive call in turn with its
  delta, and take the union of each of those''. But this is in fact incorrect
  and does not produce correct incremental programs in general.}

It is possible, however, to write a delta program for $treeP$ by hand; indeed,
the following definition provides one:
\begin{displaymath}
  \Delta_{i+1}treeP(x) = p(x) \wedge \exists y. (child(x, y) \wedge
  \Delta_itreeP(y)) \wedge \neg \exists y. (child(x,y) \wedge \neg treeP_i(y))
\end{displaymath}

This is a \emph{correct} delta program (in that using it to iteratively compute
$treeP$ gives the right answer), but it is not \emph{precise} because it
derives some facts repeatedly. We will show how to construct correct delta
programs generally using a program transformation, and show how we have some
freedom to optimize within a range of possible alternatives to improve precision or ease evaluation.

Handling extended Datalog is of more than theoretical interest \textemdash{} the research
in this paper was carried out at Semmle, which
makes heavy use of a commercial Datalog implementation
\citep{semmleWebsite,avgustinov2016ql,sereni2008adding,schafer2010type}.
Semmle's implementation includes parity-stratified negation\footnote{Parity-stratified negation means that recursive calls must
  appear under an even number of negations. This ensures that the rule remains
  monotone, so the least fixpoint still exists.},
recursive aggregates \citep{demoor2013aggregates}, and other non-standard
features, so we are faced with a dilemma: either abandon the new language
features, or abandon incremental computation.

We can tell a similar story about \emph{maintenance} of Datalog programs.
Maintenance means updating the results of the program when its inputs change,
for example, updating the value of $tc$ given a change to $e$. Again, this is a
kind of incremental computation, and there are known solutions for traditional Datalog
\citep{gupta1993maintaining}, but these break down when the language is extended.

There is a piece of folkloric knowledge in the Datalog community that hints at a
solution: the semi-naive translation of a rule corresponds to the
\emph{derivative} of that rule \citep{bancilhon1986amateur,bancilhon1986naive}[section
3.2.2]. The idea of performing incremental computation using derivatives has been
studied recently by \citet{cai2014changes}, who give an account using
\emph{change structures}. They use this to provide a framework for incrementally evaluating lambda calculus programs.

However, \citeauthor{cai2014changes}'s work isn't directly applicable to Datalog: the tricky part
of Datalog's semantics are recursive definitions and the need for the
\emph{fixpoints}, so we need some additional theory to tell us how to
handle incremental evaluation and maintenance of fixpoint computations.

This paper aims to bridge that gap by providing a solid semantic foundation for the incremental
computation of Datalog, and other recursive programs, in terms of changes and
differentiable functions.

\subsection{Contributions}

We start by generalizing change structures to
\emph{change actions} (\cref{sec:changeActions}). Change actions are simpler and weaker than change
structures, while still providing enough structure to handle incremental
computation, and have fruitful
interactions with a variety of structures (\cref{sec:moreStructures},
\cref{sec:dcpos}). In addition, change actions are not dependently typed (for
more discussion of the differences, see \cref{sec:relatedChangeStructures}),
which makes them easier to implement directly in a simply-typed language without
type erasure.

We then show how change actions can be used to perform incremental evaluation and maintenance
of non-recursive program semantics, using the formula semantics of generalized Datalog as our primary
example (\cref{sec:nonRecursiveDatalog}). Moreover, the structure of the
approach is modular, and can accommodate arbitrary additional
formula constructs (\cref{sec:extensions}).

We also provide a method of incrementally computing and maintaining fixpoints
(\cref{sec:fixpoints}). We use this to perform incremental evaluation and
maintenance of \emph{recursive} program semantics, including generalized
recursive Datalog (\cref{sec:recursiveDatalog}). This provides, to the best
of our knowledge, the world's first incremental
evaluation and maintenance mechanism for Datalog that can handle negation,
disjunction, and existential quantification. 

Finally, we provide some empirical evaluation of the effects
of our changes (\cref{sec:empirical}).

We have omitted the proofs from this paper. Most of the results have routine
proofs, but the proofs of the more substantial results
(especially those in \cref{sec:fixpoints}) are included in an appendix.

\section{Change actions and derivatives}
\label{sec:changeActions}

Incremental computation requires understanding how values \emph{change}. For
example, we can change an integer by adding a natural to it.
Abstractly, we have a set of values (the integers), and a set of changes
(the naturals) which we can ``apply'' to a value (by addition) to get a new value.

This kind of structure is well-known \textemdash{} it is a set action. It is
also very natural to want to combine changes sequentially, and if we do this
then we find ourselves with a monoid action.\footnote{
Using monoid actions for changes gives us a reason to think that
change actions are an adequate representation of changes: any subset of $A
\rightarrow A$ which is closed under composition can be
represented as a monoid action on $A$, so we are able to capture all of these as change
actions.}

\subsection{Change actions}
\label{sec:changeAction}

\begin{defn}
  A \emph{change action} is a tuple:
  \begin{displaymath}
    \cstr{A} \defeq \cstruct{A}{\changes{A}}{\cplus_A}
  \end{displaymath}
  where $A$ is a set, $\changes{A}$ is a monoid, and $\cplus_A : A \times \changes{A} \rightarrow A$ is a monoid action on $A$.\footnote{Why not
    just work with monoid actions? The reason is that while the category of
    monoid actions and the category of change actions have the same objects, they
  have different morphisms. See \cref{sec:sacts} for further discussion.}

  We will call $A$ the base set, and $\changes{A}$ the \emph{change set} of the change
  action. We will use $\splus$ for the monoid operation of $\changes{A}$, and
  $\mzero$ for its identity element. 
  When there is no risk of confusion, we will simply write $\cplus$ for $\cplus_A$.
\end{defn}

\subsubsection{Examples}
\label{sec:examples}
A typical example of a change action is $\cstruct{A^\ast}{A^\ast}{\doubleplus}$ where $A^\ast$ is the set of finite words (or lists) of $A$. 
Here we represent changes to a word made by concatenating another word onto it. 
The changes themselves can be combined using $\doubleplus$ as the monoid operation with the empty word as the identity, 
and this is a monoid action: $(a \doubleplus b) \doubleplus c = a \doubleplus \left( b \doubleplus c \right)$.

This is a very common case: any monoid $(A, \splus, 0)$ can be seen as a change action
$\cstruct{A}{(A, \splus, 0)}{\splus}$. Many practical change actions
can be constructed in this way. In particular, for any change action $\cstruct{A}{\changes{A}}{\cplus}$,
$\cstruct{\changes{A}}{\changes{A}}{\splus}$ is also a change action. This means
that we don't have to do any extra work to talk about changes to changes \textemdash{} we can 
always take $\changes{\changes{A}} = \changes{A}$.

Three examples of change actions are of particular interest to us.
First, whenever
$L$ is a Boolean algebra, we can give it the change actions $(L, L, \vee)$ and $(L, L, \wedge)$, 
as well as a combination of these (see \cref{sec:booleanAlgebras}). Second,
the natural numbers with addition have a change action $\cstr{\NN} \defeq (\NN,
\NN, +)$, which will prove useful during inductive proofs.
 
Another interesting example of change actions is \textit{semiautomata}. A semiautomaton is a triple
$(Q, \Sigma, T)$, where $Q$ is a set of states, $\Sigma$ is a (non-empty) finite input alphabet
and $T : Q \times \Sigma \rightarrow Q$ is a transition function. 
Every semiautomaton corresponds to a change action $(Q, \Sigma^*, T^*)$ on the free monoid
over $\Sigma^*$, with $T^*$ being the free extension of $T$. Conversely, every change action $\cstr{A}$
whose change set $\changes{A}$ is freely generated by a finite set corresponds to a semiautomaton.

Other recurring examples of change actions are:
\begin{itemize}
  \item $\cstr{A}_\bot \defeq \cstruct{A}{M}{\lambda(a, \change{a}). a}$, where $M$ is any monoid,
    which we call the \emph{empty} change action on any base set, since it induces no changes at all.
  \item $\cstr{A}_\top \defeq \cstruct{A}{A \rightarrow A}{\ev}$, where $A$ is an arbitrary
    set, $A \rightarrow A$ denotes the set of all functions from $A$ into itself, considered as
    a monoid under composition and $\ev$ is the usual evaluation map. We will call this the
    ``full'' change action on $A$ since it contains every possible non-redundant change.
\end{itemize}
These are particularly relevant because they are, in a sense, the ``smallest'' and ``largest''
change actions that can be imposed on an arbitrary set $A$.

Many other notions in computer science can be understood naturally in terms of change actions,
\emph{e.g.} databases and database updates, files and diffs, Git repositories and commits, even 
video compression algorithms that encode a frame as a series of changes to the previous frame.

\subsection{Derivatives}

When we do incremental computation we are usually trying to save ourselves some
work. We have an expensive function $f: A \rightarrow B$, which we've evaluated at some point
$a$. Now we are interested in evaluating $f$ after some change $\change{a}$ to
$a$, but ideally we want to avoid actually computing $f(a \cplus
\change{a})$ directly.

A solution to this problem is a function $\derive{f}: A \times \changes{A}
\rightarrow \changes B$, which given $a$ and $\change{a}$ tells us how to change
$f(a)$ to $f(a \cplus \change{a})$. We call this a \emph{derivative} of a function.

\begin{defn}
  \label{def:derivative}
  Let $\cstr{A}$ and $\cstr{B}$ be change actions.
  A \emph{derivative} of a function $f: A \rightarrow B$ is a function $\derive{f}: A \times \changes{A} \rightarrow
  \changes{B}$ such that
  \begin{displaymath}
    f(a \cplus_A \change{a}) = f(a) \cplus_B \derive{f}(a, \change{a})
  \end{displaymath}
  A function which has a derivative is 
  \emph{differentiable}, and we will write $\cstr{A} \rightarrow \cstr{B}$ for
  the set of differentiable functions.\footnote{Note that we do not require that $\derive{f}(a,
    \change{a} \splus \change{b}) = \derive{f}(a, \change{a}) \splus \derive{f}(a
    \cplus \change{a}, \change{b})$ nor that $\derive{f}(a, \mzero) = \mzero$.
    These are natural conditions, and all the 
    derivatives we have studied also satisfy them, but none of the results on
    this paper require them to hold.}
\end{defn}

Derivatives need not be unique in general, so we will speak of ``a''
derivative. Functions into ``thin'' change
actions \textemdash{} where $a \cplus \change{a} = a \cplus \change{b}$ implies $\change{a} =
\change{b}$ \textemdash{} have unique derivatives, but many change actions are not thin.
For example, because $\{0\} \cap \{1\} = \{0\}
\cap \{2\}$, $\cstruct{\powerset{\NN}}{\powerset{\NN}}{\cap}$ is not thin.

Derivatives capture the structure of incremental computation, but there are
important operational considerations that affect whether using them for
computation actually save us
any work. As we will see in a moment (\cref{prop:minusDerivatives}), for many 
change actions we will have the option
of picking the ``worst'' derivative, which merely computes $f(a \cplus \change{a})$
directly and then works out the change that maps $f(a)$ to this new value. 
While this is formally a derivative, using it certainly does not save us any work! We will be concerned 
with both the possibility of constructing correct derivatives
(\cref{sec:booleanAlgebras} and \cref{sec:fixpoints} in particular), and also in
giving ourselves a range of derivatives to choose from so that we can soundly
optimize for operational value.

For our Datalog case study, we aim to cash out the folkloric idea that
incremental computation functions via a derivative by constructing a good derivative of
the semantics of Datalog. We will do this in stages: first the non-recursive
formula semantics (\cref{sec:nonRecursiveDatalog}); and later the full, recursive, semantics 
(\cref{sec:recursiveDatalog}).

\subsection{Useful facts about change actions and derivatives}

\subsubsection{The Chain Rule}

The derivative of a function can be computed compositionally, because derivatives satisfy the standard chain rule.

\begin{thm}[The Chain Rule]
  Let $f: \cstr{A} \rightarrow \cstr{B}$, $g: \cstr{B} \rightarrow \cstr{C}$ be differentiable functions. Then $g \circ f$ is also
  differentiable, with a derivative given by
  \begin{displaymath}
    \derive{(g \circ f)}(x, \change{x}) = \derive{g}\left(f(x), \derive{f}(x, \change{x})\right)
  \end{displaymath}
  or, in curried form
  \begin{displaymath}
    \derive{(g \circ f)}(x) = \derive{g}(f(x)) \circ \derive{f}(x)
  \end{displaymath}
\end{thm}

\subsubsection{Complete change actions and minus operators}

Complete change actions are an important class of change actions, because they
have changes between \emph{any} two values in the base set.

\begin{defn}
  A change action is \emph{complete} if for any $a, b \in A$, there is
  a change $\change{a} \in \changes{A}$ such that $a \cplus \change{a} = b$.
\end{defn}

Complete change actions have convenient ``minus operators'' that allow us to
compute the difference between two values.

\begin{defn}
  A \emph{minus operator} is a function $\cminus: A \times A \rightarrow
  \changes{A}$ such that $a \cplus (b \cminus a) = b$ for all $a, b \in A$.
\end{defn}

\begin{prop}
  Let $\cstr{A}$ be a change action. Then the following are equivalent:
  \begin{itemize}
    \item $\cstr{A}$ is complete.
    \item There is a minus operator on $\cstr{A}$.
    \item For any change action $\cstr{B}$ all functions $f: {B} \rightarrow {A}$ are differentiable.
  \end{itemize}
\end{prop}

This last property is of the utmost importance, since we are often concerned with the differentiability
of functions.

\begin{prop}
  \label{prop:minusDerivatives}
  Given a minus operator $\cminus$, and a function $f$, let
  \begin{displaymath}
    \derive{f}_\cminus(a, \change{a}) \defeq f(a \cplus \change{a}) \cminus f(a)
  \end{displaymath}
  Then $\derive{f}_\cminus$ is a derivative for $f$.
\end{prop}

\subsubsection{Products and sums}
\label{sec:prodsum}

Given change actions on sets $A$ and $B$, the question immediately arises of whether there are
change actions on their Cartesian product $A \times B$ or disjoint union $A + B$. While there are
many candidates, there is a clear ``natural'' choice for both.

\begin{prop}[name=Products, restate=products]
  \label{prop:products}
  Let $\cstr{A} = \cstruct{A}{\changes{A}}{\cplus_A}$ and $\cstr{B} =
  \cstruct{B}{\changes{B}}{\cplus_B}$ be change actions.

  Then $\cstr{A} \times \cstr{B} \defeq \cstruct{A \times B}{\changes{A} \times \changes{B}}{\cplus_{\times}}$ is a change action,
  where $\cplus_{\times}$ is defined by:
  \begin{align*}
    (a, b) \cplus_{A \times B} (\change{a}, \change{b}) \defeq (a \cplus_A \change{a}, b \cplus_B \change{b})
  \end{align*}
  
  The projection maps $\pi_1$,$\pi_2$ are differentiable with respect to it.
  Furthermore, a function 
  $f : A \times B \into C$ is differentiable from $\cstr{A} \times \cstr{B}$ into $\cstr{C}$ if
  and only if, for every fixed $a \in A$ and $b \in B$, the partially applied functions 
  \begin{align*}
    f(a, \cdot) : B \into C\\
    f(\cdot, b) : A \into C
  \end{align*}
  are differentiable.
\end{prop}

Whenever $f : A \times B \rightarrow C$ is differentiable, we will sometimes use $\partial_1 f$ and
$\partial_2 f$ to refer to derivatives of the partially applied versions, i.e. if
$f'_a : B \times \changes{B} \rightarrow \changes{C}$ and
$f'_b : A \times \changes{A} \rightarrow \changes{C}$ refer to derivatives for 
$f(a, \cdot), f(\cdot, b)$ respectively, then
\begin{gather*}
  \partial_1 f : A \times \changes{A} \times B \rightarrow \changes{C}\\
  \partial_1 f(a, \change{a}, b) \defeq f'_b(a, \change{a})\\
  \partial_2 f : A \times B \times \changes{B} \rightarrow \changes{C}\\
  \partial_2 f(a, b, \change{b}) \defeq f'_a(b, \change{b})
\end{gather*}

\ifproofs
\begin{proof}
  See \cref{prf:products}.
\end{proof}
\fi

\begin{prop}[name=Disjoint unions, restate=disjointUnions]
  \label{prop:disjointUnions}
  Let $\cstr{A} = \cstruct{A}{\changes{A}}{\cplus_A}$ and $\cstr{B} =
  \cstruct{B}{\changes{B}}{\cplus_B}$ be change actions.

  Then $\cstr{A} + \cstr{B} \defeq \cstruct{A + B}{\changes{A} \times
  \changes{B}}{\cplus_{+}}$ is a change action, where $\cplus_{+}$ is defined as:
  \begin{align*}
    \iota_1 a \cplus_{+} (\change{a}, \change{b}) &\defeq \iota_1 (a \cplus_A \change{a})\\
    \iota_2 b \cplus_{+} (\change{a}, \change{b}) &\defeq \iota_2 (b \cplus_B \change{b})
  \end{align*}
  
  The injection maps $\iota_1, \iota_2$ are differentiable with respect to $\cstr{A} + \cstr{B}$. Furthermore,
  whenever $\cstr{C}$ is a change action and $f : A \rightarrow C, g: B \rightarrow C$ are differentiable,
  then so is $\left[ f, g \right]$.
\end{prop}
\ifproofs
\begin{proof}
  See \cref{prf:disjointUnions}.
\end{proof}
\fi

\subsection{Comparing change actions}

Much like topological spaces, we can compare change actions on the same
base set according to coarseness. This 
is useful since differentiability of functions between change actions is characterized
entirely by the coarseness of the actions.

\begin{defn}
  Let $\cstr{A}_1$ and $ \cstr{A}_2$ be 
  change actions on $A$. We say that $\cstr{A}_1$ is coarser than $\cstr{A}_2$ (or that $\cstr{A}_2$ is finer
  than $\cstr{A}_1$) whenever for every $x \in A$ and change $\change{a}_1 \in
  \changes{A}_1$, there is a change $\change{a}_2 \in
  \changes{A}_2$ such that $x \cplus_{A_1} \change{a}_1 = x \cplus_{A_2} \change{a}_2$.
  
  We will write $\cstr{A}_1 \leq \cstr{A}_2$ whenever $\cstr{A}_1$ is coarser than $\cstr{A}_2$.
  If $\cstr{A}_1$ is both finer and coarser than $\cstr{A}_2$, we will say that $\cstr{A}_1$
  and $\cstr{A}_2$ are equivalent.
\end{defn}

The relation $\leq$  defines a preorder (but not a partial order) on the set of all change actions 
over a fixed set A. Least and greatest elements do exist up to equivalence, and correspond
respectively to the empty change action $\cstr{A}_\bot$ and any complete change
action, such as the full change action $\cstr{A}_\top$,
defined in \cref{sec:changeAction}.

\begin{prop}
  Let $\cstr{A}_2 \leq \cstr{A}_1$, $\cstr{B}_1 \leq \cstr{B}_2$ be change actions, and suppose
  the function $f : A \rightarrow B$ is differentiable as a function from $\cstr{A}_1$ into
  $\cstr{B}_1$. Then $f$ is differentiable as a function from $\cstr{A}_2$ into $\cstr{B}_2$.
\end{prop}

A consequence of this fact is that whenever two change actions are equivalent
they can be used interchangeably without affecting which functions are differentiable. One last parallel with topology
is the following result, which establishes a simple criterion for when a change action is coarser than
another:

\begin{prop}
  Let $\cstr{A}_1, \cstr{A}_2$ be change actions on $A$. Then $\cstr{A}_1$ is coarser than $\cstr{A}_2$
  if and only if the identity function $\id : A \rightarrow A$ is differentiable from $\cstr{A}_1$ to
  $\cstr{A}_2$.
\end{prop}

\subsection{Change kernels and kernel orderings}

Consider the full change action $\cstr{\ZZ}_\top$ on the integers.
If we want to find a change that turns $99$ into $100$, there
are a few options available: for example, we can pick the function $\lambda x . 100$, or
the function $\lambda x . x + 1$. In a sense, the first one does ``too much''
work, in that it captures all the information about the target value instead of
just the difference between them.

This intuition that a change may do too much can be captured by the following definition:
\begin{defn}
  Let $\cstr{A}$ be a change action and $\change{a} \in \changes{A}$. Then the
  \emph{kernel} of $\change{a}$ is the equivalence relation $\kernel_{\change{a}}$ on $A$
  defined by
  \begin{displaymath}
    a \kernel_{\change{a}} b \iff 
     a \cplus \change{a} = b \cplus \change{a}
  \end{displaymath}
  The \emph{kernel ordering} on $\changes{A}$ is defined by
  \begin{displaymath}
    {\change{a} \kernelOrder \change{b}} \iff {{\kernel_{\change{a}}} \subseteq {\kernel_{\change{b}}}}
  \end{displaymath}
\end{defn}

We can use the kernel order, extended pointwise, to compare the different derivatives of a function.
As we will see later (\cref{sec:booleanAlgebras}), minimal derivatives under
this order sometimes correspond to ``precise'' derivatives, in the sense of derivatives that do not produce
excessively large changes.

\section{Posets and Boolean algebras}
\label{sec:moreStructures}

The semantic domain of Datalog is a complete Boolean algebra, and so our next
step is to construct a good change action for Boolean algebras. Along the way, we
will consider change actions over posets, which give us the ability to
\emph{approximate} derivatives, which will turn out to be very important in practice.

\subsection{Posets}

Ordered sets give us a constrained class of functions: monotone
functions. We can define \emph{ordered} change actions, which are those that
are well-behaved with respect to the order on the underlying set.
\footnote{If we were giving a presentation that was
generic in the base category, then this would simply be the definition of being
a change action in the category of posets and monotone maps.}

\begin{defn}
  A change action $\cstr{A}$ is \emph{ordered} if
  \begin{itemize}
    \item $A$ and $\changes{A}$ are posets.
    \item $\cplus$ is monotone as a map from $A \times \changes{A} \rightarrow A$
    \item $\splus$ is monotone as a map from $\changes{A} \times \changes{A} \rightarrow \changes{A}$
  \end{itemize}
\end{defn}

In fact, any change action whose base set is a poset induces a particularly convenient partial order
on the corresponding change set:

\begin{defn}
  $\change{a} \changeOrder \change{b}$ iff for all $a \in A$ it is the case that
  $a \cplus \change{a} \leq a \cplus \change{b}$.
\end{defn}

\begin{prop}
  Let $\cstr{A}$ be a change action on a set $A$ equipped with a partial order $\leq$ such that
  $\cplus$ is monotone in its first argument. Then $\cstr{A}$ is an ordered change action when
  $\changes{A}$ is equipped with the partial order $\changeOrder$.
\end{prop}

In what follows, we will extend the partial order $\changeOrder$ on some change
set $\changes{B}$ pointwise to functions from some $A$ into $\changes{B}$. This pointwise
order interacts nicely with derivatives, in that it gives us the following lemma:

\begin{thm}[Sandwich lemma]
  \label{thm:sandwich}
  Let $\cstr{A}$ be an change action, and $\cstr{B}$ be an ordered change action, and given functions
  $f: {A} \rightarrow {B}$ and $g: A \times \changes{A} \rightarrow
  \changes{B}$. If $\supderive{f}$ and $\subderive{f}$ are
  derivatives for $f$ such that
  \begin{displaymath}
    \supderive{f} \changeOrder g \changeOrder \subderive{f}
  \end{displaymath}
  then $g$ is a derivative for $f$.
\end{thm}

If unique minimal and maximal derivatives exist, then this gives us a 
characterisation of all the derivatives for a function.

\begin{thm}
\label{thm:derivativeCharacterization}
  Let $\cstr{A}$ and $\cstr{B}$ be change actions, with $\cstr{B}$ ordered, and given a function
  $f: {A} \rightarrow {B}$. If there exist $\subderiveM{f}$ and
  $\supderiveM{f}$ which are unique minimal and maximal derivatives of $f$,
  respectively, then the derivatives of $f$ are precisely
  the functions $\derive{f}$ such that
  \begin{displaymath}
    \subderiveM{f} \changeOrder \derive{f} \changeOrder \supderiveM{f}
  \end{displaymath}
\end{thm}
\ifproofs
\begin{proof}
  Follows easily from \cref{thm:sandwich} and minimality/maximality.
\end{proof}
\fi

This theorem gives us the leeway that we need when trying to pick a derivative: we can pick out the
bounds, and that tells us how much ``wiggle room'' we have above and below.

\subsection{Boolean algebras}
\label{sec:booleanAlgebras}

Complete Boolean algebras are a particularly nice domain for change actions
because they have a negation operator. This is very helpful for computing
differences, and indeed Boolean algebras have a complete change action.

\begin{prop}[name=Boolean algebra change actions, restate=lsuperpose]
 Let $L$ be a complete Boolean algebra. Define
  \begin{displaymath}
    \cstr{L}_\superpose \defeq \cstruct{L}{L \disjointTimes L}{\twist}
  \end{displaymath}
  where
  \begin{align*}
    L \disjointTimes L &\defeq \{ (a, b) \in L \times L \mid a \wedge b = \bot \}\\
    a \twist (p, q) &\defeq (a \vee p) \wedge \neg q
  \end{align*}
  \begin{displaymath}
    (p, q) \splus (r, s) \defeq ((p \wedge \neg s) \vee r, (q \wedge \neg r) \vee s)
  \end{displaymath}
  with identity element $(\bot, \bot)$.

  Then $\cstr{L}_\superpose$ is a complete change action on $L$.
\end{prop}
\ifproofs
\begin{proof}
  See \cref{prf:lsuperpose}.
\end{proof}
\fi

We can think of $\cstr{L}_\superpose$ as tracking changes as pairs of ``upwards'' and
``downwards'' changes, where the monoid action simply applies one after the
other, with an adjustment to make sure that the components remain disjoint.\footnote{
  The intuition that $\cstr{L}_\superpose$ is made up of an ``upwards''
  and a ``downwards'' change action glued together can in fact be made precise, but the specifics
  are outside the scope of this paper.} For example, in the powerset Boolean
algebra $\mathcal{P}(\NN)$, a change to $\{ 1, 2 \}$
might consist of \emph{adding} $\{ 3 \}$ and \emph{removing} $\{ 1 \}$,
producing $\{ 2, 3 \}$. In $\mathcal{P}(\NN)_\superpose$ this would be
represented as $(\{ 1, 2 \}) \cplus
(\{ 3 \}, \{ 1 \}) = \{ 2, 3 \}$.

Boolean algebras also have unique maximal and minimal
derivatives, under the usual partial order based on implication.\footnote{The change
set is, as usual, given the change partial order, which in this case corresponds to
the natural order on $L \times L^{\textrm{op}}$.}

\begin{prop}
  \label{prop:minimalMaximalDerivatives}
  Let $L$ be a complete Boolean algebra with the $\cstr{L}_\superpose$ change action, and
  $f: A \rightarrow L$ be a function.
  Then, the following are minus operators:
  \begin{align*}
    a \cminus_\bot b &= (a \wedge \neg b, \neg a)\\
    a \cminus_\top b &= (a, b \wedge \neg a)
  \end{align*}
  Additionally, $\derive{f}_{\cminus_{\bot}}$ and $\derive{f}_{\cminus_{\top}}$ 
  define unique least and greatest derivatives for $f$.
\end{prop}

\Cref{thm:derivativeCharacterization} then gives us bounds for
all the derivatives on Boolean algebras:

\begin{corollary}
\label{cor:booleanCharacterization}
  Let $L$ be a complete Boolean algebra with the corresponding change action
  $\cstr{L}_\superpose$, $\cstr{A}$ be an arbitrary change action, and $f: A \rightarrow
  L$ be a function. Then the derivatives of $f$ are precisely those functions
  $\derive{f}: A \times \changes{A} \rightarrow \changes{A}$ such that
  \begin{displaymath}
    \derive{f}_{\cminus_{\bot}}
    \changeOrder
    \derive{f}
    \changeOrder
    \derive{f}_{\cminus_{\top}}
  \end{displaymath}
\end{corollary}

This makes \cref{thm:derivativeCharacterization} actually usable in practice, since
we have concrete definitions for our bounds (which we will make use of in \cref{sec:datalogDifferentiability}).

Moreover, we have a nice definition of the kernel ordering, which gives us a
minus operator with a corresponding \emph{precise} derivative.

\begin{defn}
  The \emph{pointwise} order $\pointwiseOrder$ is the product order on $L \times
  L$.
\end{defn}

\begin{prop}[name=Boolean algebra kernels, restate=booleanAlgebraKernels]
  \label{prop:booleanAlgebraKernels}
  Let $L$ be a (complete) Boolean algebra with the $\cstr{L}_\superpose$ change action,
  $f: \cstr{A} \rightarrow \cstr{L}$ a (differentiable) function, and suppose 
  $\derive{f}_1$ and $\derive{f}_2$ are derivatives for $f$.

  Then $\derive{f}_1 \kernelOrder \derive{f}_2$ (extending the kernel order pointwise to derivatives)
  iff for all $a \in A$ and $\change{a} \in \changes{A}$
  \begin{align*}
    \derive{f}_1(a, \change{a}) \pointwiseOrder \derive{f}_2(a, \change{a})
  \end{align*}

  Additionally, the following is a minus operator, and $\derive{f}_{\cminus_\kernel}$ is a minimal
  derivative with respect to $\pointwiseOrder$ and $\kernelOrder$:
  \begin{displaymath}
    a \cminus_\kernel b = (a \wedge \neg b, b \wedge \neg a)
  \end{displaymath}
\end{prop}
\ifproofs
\begin{proof}
  See \cref{prf:booleanAlgebraKernels}.
\end{proof}
\fi

This shows us in what sense minimality with respect to the kernel ordering gives
us precision: the precise derivative has components which are as small as
possible (since it is also minimal with respect to $\pointwiseOrder$).
When we are working with sets or relations, this is important since it
corresponds to the actual size of components of the derivative.

\section{Derivatives for non-recursive Datalog}
\label{sec:nonRecursiveDatalog}

We now want to apply the theory we have developed to the specific case of the semantics
of Datalog. Giving a differentiable semantics for Datalog will
lead us to a strategy for performing incremental evaluation and maintenance of Datalog programs. 
To begin with, we will restrict ourselves to the non-recursive fragment of the
language \textemdash{} the formulae that make up the right hand sides of Datalog
rules. We will tackle the full program semantics in a later section, once we
know how to handle fixpoints (\cref{sec:recursiveDatalog}).

Although the techniques we are using should work for any language, Datalog
provides a non-trivial case study where the need for incremental computation is
real and pressing, as we saw in \cref{sec:intro}.

\subsection{Semantics of Datalog formulae}

Datalog is usually given a logical semantics where formulae are interpreted as first-order
logic predicates and the semantics of a program is the set of models of its constituent
predicates. We will instead give a simple denotational semantics (as is
typical when working with fixpoints \citep[See e.g.][]{compton1994stratified}) that treats a Datalog
formula as directly denoting a relation, i.e.
a set of named tuples, with variables ranging over a finite schema.

\begin{defn}
  A \emph{schema} $\Gamma$ is a finite set of names. A \emph{named tuple} over $\Gamma$ is an assignment
  of a value $v_i$ for each name $x_i$ in $\Gamma$. Given disjoint schemata 
  $\Gamma = \{ x_1, \ldots, x_n \}$ and $\Sigma = \{ y_1, \ldots, y_m \}$,
  the \emph{selection function} $\sigma_\Gamma$ is defined as
  \begin{displaymath}
    \sigma_\Gamma(\{x_1 \mapsto v_1, \ldots, x_n \mapsto v_n, y_1 \mapsto w_1, \ldots, y_m \mapsto w_m \})
    \defeq \{ x_1 \mapsto v_1, \ldots, x_n \mapsto v_n \}
  \end{displaymath}
  I.e. $\sigma_\Gamma$ restricts a named tuple over $\Gamma \cup \Sigma$ into a tuple over $\Gamma$
  with the same values for the names in $\Gamma$.
  We denote the elementwise extension of $\sigma_\Gamma$ to sets of tuples also as $\sigma_\Gamma$.
\end{defn}

We will adopt the usual closed-world assumption to give a denotation to negation.

\begin{defn}
  For any schema $\Gamma$,
  there exists a universal relation $\universalRel_\Gamma$.
  Negation on relations can then be defined as
  \begin{displaymath}
    \neg R \defeq \universalRel_\Gamma \setminus R
  \end{displaymath}
\end{defn}

This makes $\Rel_\Gamma$, the set of all subsets of $\universalRel_\Gamma$,
into a complete Boolean algebra.

\begin{figure}
  \fbox{
    \begin{minipage}[t]{0.9\textwidth}
      \vspace{-5pt}
      \begin{minipage}[t]{.45\textwidth}
      \begin{align*}
        \sem{\top}_\Gamma(\semR) &\defeq \universalRel_\Gamma\\
        \sem{\bot}_\Gamma(\semR) &\defeq \emptyset\\
        \sem{R_j}_\Gamma(\semR) &\defeq \semR_j\\
      \end{align*}
    \end{minipage}\hfill\noindent
    \begin{minipage}[t]{.45\textwidth}
      \begin{align*}
        \sem{T \wedge U}_\Gamma(\semR) &\defeq \sem{T}_\Gamma(\semR) \cap \sem{U}_\Gamma(\semR)\\
        \sem{T \vee U}_\Gamma(\semR) &\defeq \sem{T}_\Gamma(\semR) \cup \sem{U}_\Gamma(\semR)\\
        \sem{\neg T}_\Gamma(\semR) &\defeq \neg \sem{T}_\Gamma(\semR)\\
      \end{align*}
    \end{minipage}
    \vspace{-12pt}
    \begin{align*}
      \sem{\exists x . T}_\Gamma(\semR) &\defeq \sigma_\Gamma(\sem{T}_{\Gamma \cup \{ x \}}(\semR))
    \end{align*}
    \end{minipage}
  }
  \caption{Formula semantics for Datalog}
  \label{fig:datalogSemantics}
  \vspace{-12pt}
\end{figure}

\begin{defn}
  A Datalog formula $T$ whose free term variables are contained in $\Gamma$ denotes a function from 
  $\Rel_\Gamma^n$ to $\Rel_\Gamma$.
  \begin{displaymath}
    \denote{\_}_\Gamma : \Formula \rightarrow \Rel_\Gamma^n \rightarrow \Rel_\Gamma
  \end{displaymath}
  If $\semR = (\semR_1, \ldots, \semR_n)$ is a choice of a relation $\semR_i$ for each of the variables $R_i$,
  $\denote{T}(\semR)$ is inductively defined according to the rules in \cref{fig:datalogSemantics}.
\end{defn}

Since $\Rel_\Gamma$ is a complete Boolean algebra, and so is $\Rel_\Gamma^n$, $\denote{T}_\Gamma$ is
a function between complete Boolean algebras. For brevity, we will often leave the schema implicit,
as it is clear from the context.

\subsection{Differentiability of Datalog formula semantics}
\label{sec:datalogDifferentiability}

In order to actually perform our incremental computation, we first need to provide a concrete derivative for the semantics
of Datalog formulae. Of course, since $\denote{T}_\Gamma$ is a function between the complete Boolean algebras 
$\Rel_\Gamma^n$ and
$\Rel_\Gamma$, and we know that the corresponding change actions 
$\widehat{\Rel_\Gamma^n}_\superpose$ and $\widehat{\Rel_\Gamma}_\superpose$
are complete, this guarantees the existence of a derivative for $\denote{T}$.

Unfortunately, this does not necessarily provide us with an \emph{efficient} 
derivative for $\denote{T}$. The precise derivative that we know how to compute
(\cref{prop:booleanAlgebraKernels}) relies on
a minus operator:
\begin{displaymath}
  \derive{f}_{\cminus_\kernel}(a, \change{a}) = f(a \cplus \change{a}) \cminus_\kernel f(a)
\end{displaymath}

Naively computed, this expression requires computing $f(a \cplus \change{a})$
itself, which is the very thing we were trying to avoid computing!

Of course, given a concrete definition of $\cminus_\kernel$ we can simplify this
expression and hopefully make it easier to compute. But we also know from
\cref{cor:booleanCharacterization} that \emph{any} function bounded by
$\derive{f}_{\cminus_\bot}$ and $\derive{f}_{\cminus_\top}$ is a valid
derivative ($\derive{f}_{\cminus_\kernel}$ lies roughly in the middle of them),
and we can therefore optimize anywhere within that range to make a 
trade-off between ease of computation and precision.
The idea of using an approximation
to the precise derivative, and a soundness condition, appears in \citet{bancilhon1986amateur}.

There is also the question of how to compute the derivative. Since the change
set for $\widehat{\Rel}_\superpose$ is a subset of $\Rel \times \Rel$, it
is possible and indeed very natural to compute the two components via a pair of
Datalog formulae, which allows us to reuse an existing Datalog formula
evaluator. Indeed, if this process is occurring in an optimizing compiler,
the derivative formulae can themselves be optimized. This is very 
beneficial in practice, since the initial formulae may be quite complex.

This does give us additional constraints that the derivative formulae must satisfy:
for example, we need to be able to evaluate them; and we may wish to pick formulae that will be easy or cheap
for our evaluation engine to compute, even if they compute a less precise derivative.

The upshot of these considerations is that the optimal choice of derivatives is likely
to be quite dependent on the precise variant of Datalog being evaluated, and the
specifics of the evaluation engine. Here is one possibility, which is the one used at Semmle.

\subsubsection{A concrete Datalog formula derivative}
\newcommand{\bothdiff}{{\mathsf X}} 

\begin{figure}[t]
  \fbox{
    \begin{minipage}[t]{0.9\textwidth}
    \vspace{-5pt}
    \begin{minipage}[t]{.45\textwidth}
      \begin{align*}
        \updiff(\bot) &\defeq \bot\\
        \updiff(\top) &\defeq \bot\\
        \updiff(R_j) &\defeq \updiff R_j \\
        \updiff(T\vee U) &\defeq \updiff(T) \vee \updiff (U) &(\dagger)\\
        \updiff(T\wedge U) &\defeq (\updiff(T)\wedge \bothdiff(U))\\
                            & \vee (\updiff(U) \wedge \bothdiff(T))\\
        \updiff(\neg T) &\defeq \downdiff(T)\\
        \updiff(\exists x.T) &\defeq \exists x.\updiff(T) &(\dagger)
      \end{align*}
    \end{minipage}\hfill\noindent
    \begin{minipage}[t]{.45\textwidth}
      \begin{align*}
        \downdiff(\bot) &\defeq \bot\\
        \downdiff(\top) &\defeq \bot\\
        \downdiff(R_j) &\defeq \downdiff R_j \\
        \downdiff(T\vee U) &\defeq (\downdiff(T) \wedge \neg \bothdiff(U))\\
                              & \vee (\downdiff(U) \wedge \neg \bothdiff(T))\\
        \downdiff(T\wedge U) &\defeq (\downdiff(T)\wedge U) \vee (T \wedge \downdiff(U))\\
        \downdiff(\neg T) &\defeq \updiff(T)\\
        \downdiff(\exists x.T) &\defeq \exists x.\downdiff(T) \wedge \neg \exists x.\bothdiff(T)
      \end{align*}
    \end{minipage}
    \vspace{6pt}
    \begin{align*}
      \bothdiff(X) \defeq X \twist (\updiff(X), \downdiff(X))
    \end{align*}
    \end{minipage}
  }
  \caption{Upwards and downwards formula derivatives for Datalog}
  \label{fig:datalogDerivatives}
  \vspace{-12pt}
\end{figure}

In~\cref{fig:datalogDerivatives}, we define a ``symbolic'' derivative operator as a pair of mutually recursive functions,
$\updiff$ and $\downdiff$, which turn a Datalog formula $T$ into new formulae that compute
the upwards and downwards parts of the derivative, respectively. 
Our definition uses an auxiliary function, $\bothdiff$, which computes the ``neXt'' value of a term by applying the upwards and downwards derivatives.
As we expect
from a derivative, the new formulae will have additional free relation variables
for the upwards and downwards derivatives of the free relation variables of $T$.\footnote{
While the definitions usually exhibit the dualities we would expect
  between corresponding operators, there are a few asymmetries to explain.
  The asymmetry between the cases for $\updiff(T \vee U)$ and
  $\downdiff(T \wedge U)$ is for operational reasons. The symmetrical version of
  $\updiff(T \vee U)$ is $(\updiff(T) \wedge \neg U) \vee (\updiff(U) \wedge \neg
  T)$ (which is also precise). The reason we omit the negated conjuncts is simply
  that they are costly to compute and not especially helpful to our evaluation engine.
  The asymmetry between the cases for $\exists$ is because our
  dialect of Datalog does not have a primitive universal quantifier.
  If we did have one, the cases for $\exists$ would be dual to the corresponding
  cases for $\forall$.}

\newcommand{\bothchanges}{\rho}
\begin{thm}[name=Concrete Datalog formula derivatives, restate=concreteDatalog]
\label{thm:concreteDatalog}
  Let $\updiff, \downdiff, \bothdiff : \Formula \rightarrow \Formula$ be mutually recursive functions
  defined by structural induction as in \cref{fig:datalogDerivatives}.

  Then $\updiff(T)$ and $\downdiff(T)$ are disjoint, and for any schema $\Gamma$
  and any Datalog formula $T$ whose free term variables are contained in $\Gamma$,
  $\derive{\denote{T}_\Gamma} \defeq (\denote{\updiff(T)}_\Gamma, \denote{\downdiff(T)}_\Gamma)$
  is a derivative for $\denote{T}_\Gamma$.
\end{thm}
\ifproofs
\begin{proof}
  See \cref{prf:concreteDatalog}
\end{proof}
\fi

\subsubsection{Examples}

We can now give a derivative for our $treeP$ predicate by mechanically applying
the recursive functions defined in \cref{fig:datalogDerivatives}. This
derivative is composed of an upward part $\updiff(treeP(x))$ and a downward part
$\downdiff(treeP(x))$.

\begin{align*}
  &\updiff(treeP(x))\\
  &= \updiff(p(x) \wedge \neg \exists y. (child(x, y) \wedge \neg treeP(y)))\\
  &= p(x) \wedge \updiff(\neg \exists y. (child(x, y) \wedge \neg treeP(y)))\\
  &= p(x) \wedge \downdiff(\exists y. (child(x, y) \wedge \neg treeP(y)))\\
  &= p(x) \wedge \exists y. \downdiff(child(x, y) \wedge \neg treeP(y))\\
  & \quad\wedge \neg \exists y. \bothdiff(child(x,y) \wedge \neg treeP(y))\\
  &= p(x) \wedge \exists y. (child(x, y)\\
  & \quad\wedge \downdiff(\neg treeP(y))) \wedge \neg \exists y. (child(x,y) \wedge \neg \bothdiff(treeP(y)))\\
  &= p(x) \wedge \exists y. (child(x, y)\\
  & \quad\wedge \updiff(treeP(y))) \wedge \neg \exists y. (child(x,y) \wedge \neg \bothdiff(treeP(y)))\\
\end{align*}
\begin{align*} 
  &\downdiff(treeP(x))\\
  &= \downdiff(p(x) \wedge \neg \exists y. (child(x, y) \wedge \neg treeP(y)))\\
  &= p(x) \wedge \downdiff(\neg \exists y. (child(x, y) \wedge \neg treeP(y)))\\
  &= p(x) \wedge \updiff(\exists y. (child(x, y) \wedge \neg treeP(y)))\\
  &= p(x) \wedge \exists y. \updiff(child(x, y) \wedge \neg treeP(y))\\
  &= p(x) \wedge \exists y. (child(x, y) \wedge \updiff(\neg treeP(y)))\\
  &= p(x) \wedge \exists y. (child(x, y) \wedge \downdiff(treeP(y)))
\end{align*}

The upwards difference in particular is not especially easy to compute. If we naively compute it, the
third conjunct requires us to recompute the whole of the recursive part. However,
the second conjunct gives us a
guard: we only need to evaluate the third conjunct if the second conjunct is
non-empty, i.e there is \emph{some} change in the body of the existential.

This shows that our derivatives aren't a panacea: it is simply \emph{hard} to compute
downwards differences for $\exists$ (and, equivalently, upwards differences for
$\forall$) because we must check that there is no other way of deriving the same
facts.\footnote{The ``support'' data structures introduced by
  \citep{gupta1993maintaining} are an attempt to avoid this issue by
  tracking the number of derivations of each tuple.} However, we can still avoid
the re-evaluation in many cases, and the inefficiency is local to this subformula.

\subsubsection{Precision}

In practice, while the derivative given in \cref{thm:concreteDatalog} is not
precise, most of the cases are \emph{preciseness-preserving}, in that if the
subsidiary recursive cases are precise, then so is that case. The only cases
which lose precision are labelled with $\dagger$.

\subsection{Extensions to Datalog}
\label{sec:extensions}

Our formulation of Datalog formula semantics and derivatives is 
generic and modular, so it is easy to extend the language with new
formula constructs: all we need to do is add cases for $\updiff$ and $\downdiff$.

In fact, because we are using a complete change action, we can \emph{always} do this by using the maximal or
minimal derivative. This justifies our claim that we can support
\emph{arbitrary} additional formula constructs: although the maximal and minimal
derivatives are likely to be impractical, having them
available as options means that we will never be completely stymied.

This is important in practice: here is a real example from Semmle's variant of
Datalog. This includes a kind of aggregates which have well-defined recursive
semantics. Aggregates have the form
\begin{displaymath}
  r = \mathrm{agg}(p)(vs \mid T \mid U)
\end{displaymath}
where $\mathrm{agg}$ refers to an aggregation function (such as ``sum'' or
``min''), $vs$ is a sequence of variables, $p$ and $r$ are variables,
$T$ is a formula possibly mentioning $vs$, and $U$ is a formula
possibly mentioning $vs$ and $p$. The full details can been found in
\citet{demoor2013aggregates}, but for example this allows us to write
\begin{align*}
  height(n, h) \leftarrow& \neg \exists c. (child(n, c)) \wedge h = 0\\
  &\vee \exists h'. (h' = \mathrm{max}(p)(c \mid child(n, c) \mid height(c, p)) \wedge h = h' + 1)
\end{align*}
which recursively computes the height of a node in a tree.

Here is an upwards derivative for an aggregate formula:
\begin{align*}
  \updiff(r = \mathrm{agg}(p)(vs \mid T \mid U)) \defeq \exists vs. (T \wedge \updiff{U}) \wedge r = \mathrm{agg}(p)(vs \mid T \mid U)
\end{align*}

While this isn't a precise derivative, it is still substantially cheaper than
re-evaluating the whole subformula, as the first conjunct acts as a guard,
allowing us to skip the second conjunct when $U$ has not changed.


\section{Changes on functions}

So far we have found change actions for spaces that are a good representation of
simple data, but we would also like to have change actions on
\emph{function} spaces. This would enable us to have derivatives for higher-order languages, and in particular derivatives for
fixpoint operators $\fixpoint : (A \rightarrow A) \rightarrow A$, which are higher-order functions. 

Function spaces, however, differ from products and disjoint unions in that, given change actions $\cstr{A}$ and $\cstr{B}$, there is no obvious choice
of a ``best'' change action structure on $A \rightarrow B$.
Instead of settling on a concrete choice of a change action, we will instead
pick out subsets of ``well-behaved'' change actions on function spaces.

\begin{defn}[Functional Change Action]
  \label{def:functionalChanges}
  Given change actions $\cstr{A}$ and $\cstr{B}$ and a set $U \subseteq A \rightarrow B$, a change action
  $\cstr{U} = (U, \changes U, \cplus_U)$ is \emph{functional} whenever the evaluation map $\ev : U \times A \rightarrow B$
  is differentiable, that is to say, whenever there exists a function 
  $\derive{\ev} : (U \times A) \times (\changes U \times \changes A) \rightarrow \changes{B}$ such that:
  \begin{displaymath}
    (f \cplus_U \change{f})(a \cplus_A \change{a}) = 
    f(a) \cplus_B \derive{\ev}((f, a), (\change{f}, \change{a}))
  \end{displaymath}
  We will write $\cstr{U} \subseteq \cstr{A} \Rightarrow \cstr{B}$ whenever 
  $U \subseteq A \rightarrow B$ and $\cstr{U}$ is functional.
\end{defn}

We have defined functional change actions on arbitrary subsets $U \subseteq A \rightarrow B$ for two
reasons. First, it will later allow us to restrict ourselves to spaces of monotone or continuous
functions. But more importantly, functional change actions are necessarily made up of differentiable
functions, and thus a functional change action may not exist for the entire function space
$A \rightarrow B$.

\begin{prop}
  \label{prop:differentiableFunctionalChanges}
  Let $\cstr{U} \subseteq \cstr{A} \Rightarrow \cstr{B}$ be a functional change action. Then every 
  $f \in U$ is differentiable, with a derivative $\derive{f}$ given by:
  \begin{displaymath}
    \derive{f}(x, \change{x}) = \derive{\ev}((f, x), (0, \change{x}))
  \end{displaymath}
\end{prop}

\subsection{Pointwise functional change actions}
\label{sec:pointwiseFunctional}

It is, in general, hard to find functional change actions on the set
of differentiable functions $\cstr{A} \rightarrow \cstr{B}$. Fortunately, 
in many important cases there is a simple change action structure on the space of differentiable 
functions.

\begin{defn}[Pointwise functional change action]
  Let $\cstr{A}$ and $\cstr{B}$ be change actions. The \emph{pointwise functional change action} 
  $\cstr{A} \ptfunc \cstr{B}$, when it is defined,
  is given by $(\cstr{A} \rightarrow \cstr{B}, A \rightarrow \changes{B}, \cplus_\rightarrow)$, with
  the monoid structure $(A \rightarrow \changes{B}, \splus_\rightarrow, 0_\rightarrow)$ and the action 
  $\cplus_\rightarrow$ defined by:
  \begin{align*}
    (f \cplus_\to \delta f)(x) &\defeq f(x) \cplus_B \delta f (x)\\
    (\delta f \splus_\rightarrow \delta g)(x) &\defeq \delta f(x) \splus_B \delta g(x)\\
    0_\rightarrow (x) &\defeq 0_B
  \end{align*}
\end{defn}

The above definition isn't always well-typed, since given $f : \cstr{A} \rightarrow \cstr{B}$ and
$\delta f : A \rightarrow \changes{B}$ there is, in general, no guarantee that 
$f \cplus_\rightarrow \delta f$ is differentiable. We present two simple criteria that guarantee this.

\begin{thm}
  \label{thm:functionalCAct}
  Let $\cstr{A}$ and $\cstr{B}$ be change actions, and suppose that $\cstr{B}$ satisfies one of the
  following conditions:
  \begin{itemize}
    \item $\cstr{B}$ is a complete change action.
    \item The change action $\widehat {\changes{B}} \defeq (\changes{B}, \changes{B}, \splus_B)$ is 
      complete and 
      $\cplus_B : B \times \changes{B} \to B$ is differentiable.
  \end{itemize}
  Then the pointwise functional change action 
  $(\cstr{A} \rightarrow \cstr{B}, A \rightarrow \changes{B}, \cplus_\rightarrow)$ is well defined. 
  \footnote{
    Either of these conditions is enough to guarantee that the pointwise functional change action
    is well defined, but it can be the case that $\cstr{B}$ satisfies neither and yet pointwise
    change actions into $\cstr{B}$ do exist. A precise account of when pointwise functional change
    actions exist is outside the scope of this paper.
  }
\end{thm}
As a direct consequence of this, it follows that whenever $L$ is a Boolean algebra, the
pointwise functional change action $\cstr{A} \ptfunc \cstr{L}_\superpose$ is well-defined.

Pointwise functional change actions are functional in the sense of \cref{def:functionalChanges}. 
Moreover, the derivative of the evaluation map is quite easy to compute. 
\begin{prop}[Derivatives of the evaluation map]
\label{prop:evDerivatives}
  Let $\cstr{A}$ and $\cstr{B}$ be change actions such that the pointwise functional change action
  $\cstr{A} \ptfunc \cstr{B}$ is well defined, and let
  $f: \cstr{A} \rightarrow \cstr{B}$,
  $a \in A$, $\change{a} \in \changes{A}$,
  $\change{f} \in A \rightarrow \changes{B}$.

  Then, by taking a derivative of $f$ we obtain the following derivative for the evaluation map:
  \begin{displaymath}
    \derive{\ev}_1((f, a), (\change{f}, \change{a})) 
    \defeq \derive{f}(a, \change{a}) \splus \change{f}(a \cplus \change{a})
  \end{displaymath}

  Alternatively, by taking a derivative of $f \cplus \change{f}$ we can obtain another derivative
  for the evaluation map:
  \begin{displaymath}
    \derive{\ev}_2((f, a), (\change{f}, \change{a})) 
    \defeq \change{f}(a) \splus \derive{(f \cplus \change{f})}(a, \change{a})
  \end{displaymath}
\end{prop}

Having pointwise function changes allows us to actually compute a derivative of the
evaluation map as shown. In practice, this means
that we will only be able to use the results in \cref{sec:fixpoints}
(incremental computation and derivatives of fixpoints) when
we have pointwise change actions, or where we have some other way of computing
a derivative of the evaluation map. 

\section{Directed-complete partial orders and fixpoints}

Directed-complete partial orders (dcpos) equipped with a least element, are an
important class of posets. They provide us with the ability to take
\emph{fixpoints} of (Scott-)continuous maps, which is especially important for interpreting recursion in program semantics.

\subsection{Dcpos}
\label{sec:dcpos}

As before, we can define change actions on dcpos, rather than sets, as change
actions whose base and change sets are endowed with a dcpo structure, and where
the monoid operation and action are continuous.

\begin{defn}
  A change action $\cstr{A}$ is \emph{continuous} if
  \begin{itemize}
    \item $A$ and $\changes{A}$ are dcpos.
    \item $\cplus$ is Scott-continuous as a map from $A \times \changes{A} \rightarrow A$.
    \item $\splus$ is Scott-continuous as a map from $\changes{A} \times \changes{A} \rightarrow \changes{A}$.
  \end{itemize}
\end{defn}

Unlike the case for posets, the change order $\changeOrder$ does \emph{not}, in general,
induce a dcpo on $\changes{A}$. As a counterexample, consider 
the change action $(\overline{\NN}, \NN, +)$, where $\overline{\NN}$ denotes the dcpo of natural numbers
extended with positive infinity.

A key example of a continuous change action is the $\cstr{L}_\superpose$ change
action on Boolean algebras.

\begin{prop}[name=Boolean algebra continuity, restate=booleanAlgebraContinuous]
  \label{prop:booleanAlgebraContinuous}
  Let $L$ be a Boolean algebra. Then $\cstr{L}_\superpose$ is a continuous
  change action.
\end{prop}
\ifproofs
\begin{proof}
  See \cref{prf:booleanAlgebraContinuous}.
\end{proof}
\fi

For a general overview of results in domain theory and dcpos, we refer the reader to an
introductory work such as \cite{abramsky1994domain}, but we state here some specific results that
we shall be using, such as the following, whose proof can be found in 
\cite[Lemma~3.2.6]{abramsky1994domain}:

\begin{prop}
  \label{prop:distributivityLimit}
  A function $f : A \times B \rightarrow C$ is continuous iff it is continuous in each variable
  separately.
\end{prop}

It is a well-known result in standard calculus that the limit of an absolutely convergent sequence of
differentiable functions $\{f_i\}$ is itself differentiable, and its derivative is equal to the limit
of the derivatives of the $f_i$. A remarkable consequence of the previous distributivity property
is the following analogous result:

\begin{corollary}
  \label{cor:diffContinuous}
  Let $\cstr{A}$ and $\cstr{B}$ be change actions, with $\cstr{B}$ continuous and let $\{f_i\}$ and $\{\derive{f_i}\}$ be
  $I$-indexed directed families of functions in $A \rightarrow B$ and $A \times \changes{A} \rightarrow \changes{B}$ respectively.

  Then, if for every $i \in I$ it is the case that $\derive{f_i}$ is a derivative of $f_i$, then $\bigsqcup_{i \in I} \derive{f_i}$ is
  a derivative of $\bigsqcup_{i \in I} f_i $.
\end{corollary}
\ifproofs
\begin{proof}
  It suffices to apply $\cplus$ and \cref{prop:distributivityLimit} to the directed families $\{ f_i(a) \}$ and
  $\{ \derive{f_i}(a, \change{a}) \}$.
\end{proof}
\fi

We also state the following additional fixpoint lemma. This is a specialization of
Beki\'c's Theorem \citep[][section 10.1]{winskel1993formal}, but it has a straightforward direct proof.

\begin{prop}[name=Factoring of fixpoints, restate=factoringFixpoints]
  \label{prop:factoringFixpoints}
  Let $A$ and $B$ be dcpos, $f : A \rightarrow A$ and $g: A \times B \rightarrow B$ be continuous, and let
  \begin{displaymath}
    h(a, b) \defeq (f(a), g(a, b))
  \end{displaymath}
  Then
  \begin{displaymath}
    \lfp(h) = (\lfp(f), \lfp(\lambda b . g(\lfp(f), b)))
  \end{displaymath}
\end{prop}
\ifproofs
\begin{proof}
  See \cref{prf:factoringFixpoints}.
\end{proof}
\fi

\subsection{Fixpoints}
\label{sec:fixpoints}

Fixpoints appear frequently in the semantics of languages with recursion. If we
can give a generic account of how to compute fixpoints using change actions,
then this gives us a compositional way of extending a derivative for the
non-recursive semantics of a language to a derivative that can also handle recursion.
We will later apply this to full recursive Datalog (\cref{sec:datalogIncr}).

\subsubsection{Iteration functions}
\label{sec:iteration}

Over directed-complete partial orders we can define a least fixpoint operator
$\lfp$ in terms of the
iteration function $\iter$:
\begin{align*}
  &\iter : (A \rightarrow A) \times \NN \rightarrow A\\
  &\iter(f, n) \defeq f^n(\bot)\\
  &\lfp : (A \rightarrow A) \rightarrow A\\
  &\lfp(f) \defeq \bigsqcup_{n \in \NN} \iter(f, n) \tag{where $f$ is continuous}
\end{align*}

The iteration function is the basis for all the results in this section:
we can take a partial derivative with respect to $n$, and this will give us a way to get
to the next iteration incrementally; and we can take the partial derivative
with respect to $f$, and this will give us a way to get from iterating $f$ to iterating $f
\cplus \change{f}$.

\subsubsection{Incremental computation of fixpoints}

The following theorems provide a
generalization of semi-naive evaluation to any differentiable function over a
continuous change action. Throughout this section we will assume that we have a continuous change action
$\cstr{A}$, and any reference to the change action $\cstr{\NN}$ will refer to the obvious monoidal
change action on the naturals defined in \cref{sec:examples}.

Since we are trying to incrementalize the iterative step, we start by taking the partial
derivative of $\iter$ with respect to $n$.

\begin{prop}[name=Derivative of the iteration map with respect to $n$, restate=iterDerivativesN]
  \label{prop:iterDerivativesN}
  Let $\cstr{A}$ be a complete change action and let $f: A \rightarrow A$ be a differentiable function. 
  Then $\iter$ is differentiable with respect to its second
  argument, and a partial derivative is given by:
  \begin{align*}
    &\partial_2{\iter}: (A \rightarrow A) \times \NN \times \changes{\NN} \rightarrow \changes{A}\\
    &\partial_2{\iter}(f, 0, m) \defeq \iter(f, m) \cminus \bot\\
    &\partial_2{\iter}(f, n+1, m) \defeq \derive{f}(\iter(f, n), \partial_2{\iter}(f, n, m))
  \end{align*}
\end{prop}
\ifproofs
\begin{proof}
  See \cref{prf:iterDerivativesN}.
\end{proof}
\fi

By using the following recurrence relation, 
we can then compute $\partial_2{\iter}$ along with $\iter$ simultaneously:
\begin{align*}
  &\nextiter_f : A \times \changes{A} \rightarrow A \times \changes{A}\\
  &\nextiter_f (\bot, \bot) \defeq (\bot, f(\bot) \cminus \bot)\\
  &\nextiter_f (a, \change{a}) \defeq (a \cplus \change{a}, \derive{f}(a, \change{a}))
\end{align*}
Which has the property that
\begin{align*}
  &\nextiter_f^n (\bot, \bot) = (\iter(f, n), \partial_2{\iter}(f, n, 1))
\end{align*}

This gives us a way to compute a fixpoint incrementally, by adding successive
changes to an accumulator until we reach it. This is exactly how
semi-naive evaluation works, with the delta relation and the accumulator relation.

\begin{thm}[name=Incremental computation of least fixpoints, restate=fixpointIter]
\label{thm:fixpointIter}
  Let $\cstr{A}$ be a complete, continuous change action, $f: \cstr{A} \rightarrow
  \cstr{A}$ be continuous and differentiable.

  Then $\lfp(f) = \bigsqcup_{n \in \NN}(\pi_1(\nextiter_f^n(\bot, \bot)))$.\footnote{
    Note that we have \emph{not} taken the fixpoint of $\nextiter_f$, since it is
    not continuous.}
\end{thm}
\ifproofs
\begin{proof}
  See \cref{prf:fixpointIter}.
\end{proof}
\fi

\subsubsection{Derivatives of fixpoints}
\label{sec:fixpointDerivatives}

In the previous section we have shown how to use derivatives to compute fixpoints
more efficiently, but we also want to take the derivative of the fixpoint
operator itself. A typical use case for this is where we have calculated some fixpoint
\begin{displaymath}
  F_{E} \defeq \fixpoint (\lambda X . F(E, X))
\end{displaymath}
then update the parameter $E$ with some change $\change{E}$ and wish to compute the new
value of the fixpoint, i.e.
\begin{displaymath}
  F_{E \cplus \change{E}} \defeq \fixpoint (\lambda X . F(E \cplus \change{E}, X))
\end{displaymath}

If $F : B \times A \rightarrow A$ and $\cstr{A}$ is a complete change action where
$\cstr{A} \ptfunc \cstr{A}$ is well-defined, then 
\begin{displaymath}
  \lambda X . F(E \cplus \change{E}, X) = (\lambda X . F(E, X)) \cplus_{pt}
  (\lambda X . F(E \cplus \change{E}, X) \cminus F(E, X))
\end{displaymath}
i.e. we are applying a change to the \emph{function} whose fixpoint we are taking.

In Datalog this would allow us to update a recursively defined relation given an
update to one of its non-recursive dependencies, or the extensional database.
For example, we might want to take the transitive closure relation
and update it by changing the edge relation $e$.\footnote{See
  \cref{sec:workedExample} for a worked example of this process.}

However, this requires us to provide a derivative for the fixpoint operator
$\fixpoint$ with respect to the function whose fixpoint is being taken.

\begin{defn}[Derivatives of fixpoints]
\label{def:fixpointDerivatives}
  Let $\cstr{A}$ be a change action, let $\cstr{U} \subseteq \cstr{A} \Rightarrow \cstr{A}$ be a functional
  change action (not necessarily pointwise) and suppose $\fixpoint_U$ and $\fixpoint_{\changes{A}}$ are fixpoint
  operators for endofunctions on $U$ and $\changes{A}$ respectively.
  
  Then we define
  \begin{align*}
    &\adjust : U \times \changes{U} \rightarrow (\changes{A} \rightarrow \changes{A})\\
    &\adjust(f, \change{f}) \defeq \lambda\ \change{a} . \derive{\ev}((f, \fixpoint_U(f)), (\change{f}, \change{a}))\\
    &\derive{\fixpoint_U} : U \times \changes{U} \rightarrow \changes{A}\\
    &\derive{\fixpoint_U}(f, \change{f}) \defeq \fixpoint_{\changes{A}}(\adjust(f, \change{f}))
  \end{align*}
\end{defn}

The suggestively named $\derive{\fixpoint_U}$ will in fact turn out to be a
derivative \textemdash{} for \emph{least} fixpoints. The appearance of
$\derive{\ev}$, a derivative of the evaluation map, in the definition of
$\adjust$ is also no coincidence: as evaluating a fixpoint consists of many
steps of applying the evaluation map, so computing the derivative of a fixpoint
consists of many stages of applying the derivative of the evaluation
map.\footnote{Perhaps surprisingly, the authors first discovered an expanded
  version of this formula, and it was only later that we realised the remarkable
  connection to $\derive{\ev}$.}

\begin{thm}[name=Pseudo-derivatives of fixpoints, restate=fixpointPseudoDerivatives]
\label{thm:fixpointPseudoDerivatives}
  Let $\cstr{A}$ be a change action, $\cstr{U} \subseteq \cstr{A} \Rightarrow
  \cstr{A}$ be a functional change action, $\fixpoint_{U}$, and 
  $\fixpoint_{\changes A}$ be as in \cref{def:fixpointDerivatives}. Then, for any
  differentiable function $f \in U$ and any change $\change{f} \in \changes{U}$,
  $\fixpoint_U(f) \cplus \derive{\fixpoint_U}(f, \change{f})$ is a fixpoint
  of $f \cplus \change{f}$.
\end{thm}
\ifproofs
\begin{proof}
  See \cref{prf:fixpointPseudoDerivatives}.
\end{proof}
\fi

This is not enough to give us a true derivative, because we have only shown
that $\fixpoint_U(f) \cplus \derive{\fixpoint_U}(f, \change{f})$ computes \emph{a} 
fixpoint for $f \cplus \change{f}$, but not necessarily
the same one computed by $\fixpoint_U{(f \cplus \change{f})}$.

However, if we restrict ourselves to directed-complete partial orders, least
fixpoints, and continuous change actions, then $\derive{\lfp}$ (using the
derivative defined for fixpoint operators in \cref{def:fixpointDerivatives}) \emph{is} a
derivative of $\lfp$. This is not too onerous a restriction, since this is
a very natural setting for computing fixpoints.

Since $\lfp$ is characterized as the limit of a chain of functions,
\cref{cor:diffContinuous} suggests a way to compute its derivative. It suffices to find a derivative
$\derive{\iter_n}$ of each iteration map 
such that the resulting set $\{ \derive{\iter_n} \mid n \in \NN\}$ is directed, 
which will entail that $\bigsqcup_{n \in \NN}\derive{\iter_n}$ is a derivative of $\lfp$.

These correspond to the first partial derivative of $\iter$ \textemdash{} this time with respect to
$f$. While we are differentiating with respect to $f$, we are still going to
need to define our derivatives inductively in terms of $n$.

\begin{prop}[name=Derivative of the iteration map with respect to $f$, restate=iterDerivativesF]
  \label{prop:iterDerivativesF}
  $\iter$ is differentiable with respect to its first argument and a derivative is given by:
  \begin{align*}
    &\partial_1{\iter} : (A \rightarrow A) \times \changes{(A \rightarrow A)} \times \NN \rightarrow \changes{A}\\
    &\partial_1{\iter} (f, \change{f}, 0) \defeq \bot_{\changes{A}}\\
    &\partial_1{\iter} (f, \change{f}, n + 1) \defeq \derive{\ev}((f, \iter(f, n)), (\change{f}, \partial_1{\iter}(f, \change{f}, n)))
  \end{align*}
\end{prop}
\ifproofs
\begin{proof}
  See \cref{prf:iterDerivativesF}.
\end{proof}
\fi

As before, we can now compute $\partial_1{\iter}$ together with $\iter$ by
mutual recursion.\footnote{
  In fact, the recursion here is not \emph{mutual}: the first component does not
  depend on the second. However, writing it in this way makes it
  amenable to computation by fixpoint, and we will in fact be able to avoid the
  recomputation of $\iter_n$ when we show that it is equivalent to $\derive{\lfp}$.
}
\begin{align*}
  &\nextiter_{f, \change{f}} : A \times \changes{A} \rightarrow A \times \changes{A}\\
  &\nextiter_{f, \change{f}} (a, \change{a}) \defeq (f(a), \derive{\ev}((f, a), (\change{f}, \change{a})))
\end{align*}
Which has the property that
\begin{align*}
  &\nextiter_{f, \change{f}}^n (\bot, \bot) = (\iter(f, n), \partial_1{\iter}(f, \change{f}, n)).
\end{align*}

This indeed provides us with a function whose limit we can take, showing
that $\derive{\lfp}$ is a true derivative.

\begin{thm}[name=Derivatives of least fixpoint operators, restate=leastFixpointDerivatives]
  \label{thm:leastFixpointDerivatives}
  Let
  \begin{itemize}
    \item $\cstr{A}$ be a continuous change action
    \item $U$ be the set of continuous functions $f : A \rightarrow A$, 
      with a functional change action $\cstr{U} \subseteq \cstr{A} \Rightarrow \cstr{A}$
    \item $f \in U$ be a continuous, differentiable function
    \item $\change{f} \in \changes{U}$ be a function change
    \item $\derive{\ev}$ be a derivative of the evaluation map which is continuous with
      respect to $a$ and $\change{a}$.
  \end{itemize}
  Then $\derive{\lfp}$ is a derivative of $\lfp$.
\end{thm}
\ifproofs
\begin{proof}
  See \cref{prf:leastFixpointDerivatives}.
\end{proof}
\fi

Computing this derivative still requires computing a fixpoint \textemdash{} over the change
lattice \textemdash{} but this may still be significantly less expensive than
recomputing the full new fixpoint.

\section{Derivatives for recursive Datalog}
\label{sec:recursiveDatalog}

Given the non-recursive semantics for a language, we can extend it to handle
recursive definitions using fixpoints. \Cref{sec:fixpoints} lets us extend our
derivative for the non-recursive semantics to a derivative for the recursive
semantics, as well as letting us compute the fixpoints themselves
incrementally. 

Again, we will demonstrate the technique with Datalog, although the approach is generic.

\subsection{Semantics of Datalog programs}

First of all, we define the usual ``immediate consequence operator'' which
computes ``one step'' of our program semantics.

\begin{defn}
  Given a program $\mathbb{P} = (P_1, \dots, P_n)$, where $P_i$ is a predicate,
  with schema $\Gamma_i$, the \emph{immediate consequence operator} $\consq: \Rel^n \rightarrow \Rel^n$ is defined 
  as follows:
  \begin{displaymath}
    \consq(\semR_1, \dots, \semR_n) 
    = (\denote{P_1}_{\Gamma_1}(\semR_1, \dots, \semR_n), \dots, \denote{P_n}_{\Gamma_n}(\semR_1, \dots, \semR_n))
  \end{displaymath}
\end{defn}

That is, given a value for the program, we pass in all the relations
to the denotation of each predicate, to get a new tuple of relations.

\begin{defn}
  The semantics of a program $\mathbb{P}$ is defined to be
  \begin{displaymath}
    \denote{\mathbb{P}} \defeq \lfp_{\Rel^n}(\consq)
  \end{displaymath}
  and may be calculated by iterative application of $\consq$ to $\bot$ until
  fixpoint is reached.
\end{defn}

Whether or not this program semantics exists will depend on whether the fixpoint
exists. Typically this is ensured by constraining the program such that $\consq$
is monotone (or, in the context of a dcpo, continuous). We do not require monotonicity
to apply \cref{thm:fixpointIter} (and hence we can incrementally
compute fixpoints that happen to exist even though the generating function is
not monotonic), but it is required to apply \cref{thm:leastFixpointDerivatives}.

\subsection{Incremental evaluation of Datalog}
\label{sec:datalogIncr}

We can easily extend a derivative for the formula semantics to a derivative for
the immediate consequence operator $\consq$. Putting this together with the
results from \cref{sec:fixpoints}, we get our two big results.

\begin{corollary}
\label{thm:diffEval}
  Datalog program semantics can be evaluated incrementally.
\end{corollary}
\ifproofs
\begin{proof}
  Corollary of \cref{thm:fixpointIter} and \cref{corollary:consqDiff}.
\end{proof}
\fi

This is known (semi-naive evaluation), but our proof is more
modular, so we will be able to extend this result more easily.

\begin{corollary}
\label{thm:diffUpdate}
  Datalog program semantics can be incrementally maintained with changes to
  extensional (EDB) relations.
\end{corollary}
\ifproofs
\begin{proof}
  Corollary of \cref{thm:leastFixpointDerivatives} and \cref{corollary:consqDiff}.
\end{proof}
\fi

This is known \citep[see][and successors]{gupta1993maintaining},
but again, the proof is now modular so we can extend it.

\subsubsection{Worked example of updating a recursive Datalog program}
\label{sec:workedExample}

The algorithm in \cref{thm:leastFixpointDerivatives} is very abstract, and it is
hard to see how it will work out in practice. It is therefore worth doing a
simple worked example.


Consider the $tc$ program from \cref{sec:intro}:
\begin{align*}
  tc(x, y) &\leftarrow e(x, y) \vee \exists z.(e(x, z) \wedge tc(z, y))
\end{align*}

We will start with an edge relation $e_1$ and change it to a new edge relation
$e_2$ by applying a change $\change{e}$, which both adds and removes some values.
\begin{align*}
  e_1 &= \{(1,2), (2,3), (3,4), (5,6) \}\\
  \change{e} &= (\{(4,5)\}, \{(2,3)\})\\
  e_2 &= e_1 \cplus \change{e}\\
      &=\{ (1,2), (3,4), (4,5), (5,6) \}
\end{align*}

We want to update the fixpoint of $tc$ using $e_1$, which we will call
$tc_{e_1}$,  to the fixpoint of $tc$ using $e_2$, which we will call $tc_{e_2}$.
As we have seen in \cref{thm:leastFixpointDerivatives},
we can do this by computing the fixpoint $\change{w} = \lfp(\adjust(tc_{e_1}, \change{tc}))$. 

A reminder of the expanded definition of $\adjust$:
\begin{align*}
  \adjust(f, \change{f}) &= \lambda\ \change{a}. \change{f}(\lfp(f)) \splus \derive{(f \cplus \change{f})}(\lfp(f), \change{a})
\end{align*}

We've chosen the second definition of $\ev$ from \cref{prop:evDerivatives}, since
we already know what $f \cplus \change{f}$ looks like \textemdash{} it is simply
$\denote{tc_{e_2}}$ \textemdash{} and $\change{tc}(\lfp(tc_{e_1}))$ can be
computed once up front and reused throughout the computation.

We need the derivative of $tc_{e_2}$:
\begin{align*}
  \updiff(tc_{e_2}(x, y)) \leftarrow & \exists z. (e_2(x,z) \wedge \updiff(tc_{e_2}(z, y)))\\
  \downdiff(tc_{e_2}(x, y)) \leftarrow & \neg e_2(x, y)\\
    &\wedge
    \exists z . (e_2(x, z) \wedge \downdiff(tc_{e_2}(z, y)))\\
    &\wedge
    \neg \exists z . (e_2(x, z) \wedge \bothdiff(tc_{e_2}(z,y)))
\end{align*}

We also need $\change{tc}$:
\begin{align*}
  \change{tc}(f) = &\\
  &(e_2(x,y) \vee \exists z . (e_2(x,z) \wedge f(z,y))) \\
  &\cminus (e_1(x,y) \vee \exists z . (e_1(x,z) \wedge f(z,y)))
\end{align*}

We can now evaluate the fixpoint.\footnote{We would of course like to evaluate this fixpoint
  incrementally, which we can do with exactly the same theoretical machinery.}
Here is how the changes evolve:
\begin{center}
  \begin{tabular} {p{3.5em} p{10em} p{10em}}
    Iteration & Additions to $\updiff$ & Additions to $\downdiff$ \\
    \toprule
    1 & $\{ (4,5), (4,6) \}$ & $\{ (2,3), (2,4) \}$\\
    2 & $\{ (3,5), (3,6) \}$ & $\{ (1,3), (1,4) \}$\\
    3 & (as above) & (as above) \\
    \bottomrule
  \end{tabular}
\end{center}
\medskip

This results in $\change{w} = (\{ (3,5), (3,6), (4,5), (4,6)\}, \{(1,3), (1,4), (2,3), (2,4)\})$.
Applying the change and shows that we have indeed computed the new fixpoint.
\begin{align*}
  tc_{e_1} \cplus \change{w} &= \{(1,2), (1,3), (1,4), (2,3), (2,4), (5,6)\} \cplus \change{w}\\
  &= \{(1,2), (3,5), (3,6), (4,5), (4,6), (5,6)\}\\
  &= tc_{e_2}
\end{align*}

\section{Empirical evaluation}
\label{sec:empirical}

The necessity of semi-naive evaluation for practical Datalog engines is
well-established, as we have seen, it can lead to asymptotic improvements.
One might ask whether we actually gain anything by further expanding the set of predicates
which we can evaluate incrementally \textemdash{} what if we just evaluated the
predicates that we can evaluate with semi-naive with semi-naive, and evaluated
the other predicates naively? If the non-semi-naive predicates are rare enough, then perhaps this
will not matter.

We tested this empirically using Semmle's commercial Datalog engine, which uses
exactly the derivatives described in \cref{fig:datalogDerivatives}. We used two
versions of the engine:
\begin{itemize}
  \item A ``full'' version using the incremental evaluation strategy described in this paper
  \item A ``semi-naive'' version which evaluates predicates using traditional
    semi-naive evaluation, falling back to naive evaluation when it cannot be applied.
\end{itemize}

We used Semmle's profiling infrastructure, which runs over 800 queries (compiled to
Datalog programs) on a wide range
of real-world databases. The queries are compiled from Semmle's QL language
\citep{avgustinov2016ql}, which produces large Datalog programs \textemdash{}
a typical size is over 30,000 lines of Datalog. The databases represent
real-world programs that are subject to analysis, for example the Linux kernel,
which is 1.8 MLOC and produces a database of size 783Mb.

The profiler runs each query on each database with each version of the compiler
Queries are run with a (generous) timeout \textemdash{} as we will see, many
of the queries will in fact time out.

Comparing the results for the two incrementalization strategies gives
the following high-level picture:

\begin{center}
  \begin{tabular} {l l}
    Statistic & Value \\
    \toprule
    Total runs & 2308\\
    Runs which time out in ``semi-naive'' but not in ``full'' & 629 (27\% of all runs)\\
    Runs which take more time to evaluate in ``semi-naive'' than ``full'' & 1049 (45\% of all runs)\\
    Runs which take less time to evaluate in ``semi-naive'' than ``full'' & 630 (27\% of all runs)\\
    \bottomrule
  \end{tabular}
\end{center}
\medskip

This presents a somewhat inconsistent outcome, with some queries improving and some
regressing. However, the most interesting category is the category of timeouts
that appear in ``semi-naive'' but not in ``full''.
Further investigation reveals that these regress very badly indeed. For many of
these query/database pairs, the authors have been unable to allocate enough memory and time for
the query to actually complete. Ad hoc investigation reveals that many of those
have a similar evaluation pattern to the $tc$ example in \cref{sec:intro}
\textemdash{} where there is a very long chain of deductions needed to fully
evaluate the predicate. In this case naive evaluation will scan the entire
accumulator predicate in every iteration, leading to quadratic runtime. 

This shows us more-or-less what we expected: for many queries the results are
moderate and broadly positive, but for some queries incremental evaluation is
essential for evaluation to be feasible at all.

We have not performed evaluation of different variants of
\cref{fig:datalogDerivatives} against each other. Changes to the precise form of
the derivative are important, but mostly a matter of adapting them to the
particular database engine, and so of less general interest.

\section{Related work}

\subsection{Change actions and incremental computation}

\subsubsection{Change structures}
\label{sec:relatedChangeStructures}

The seminal paper in this area is \citet{cai2014changes}. We deviate from
that excellent paper in three regards: the
inclusion of minus operators, the nature of function changes, and the use of
dependent types.

We have omitted minus operators from our definition because
there are many interesting change actions that are not complete and so cannot
have a minus operator. Where we can find a change structure with a minus operator, often we are
forced to use unwieldy representations for change sets, and
\citeauthor{cai2014changes} cite this as their reason for using a dependent
type of changes. For example, the monoidal change actions on sets and lists are clearly
useful for incremental computation on streams, yet they do not admit minus
operators \textemdash{} instead, one would
be forced to work with e.g. multisets admitting negative arities, as \citeauthor{cai2014changes} do.

Our function changes (when well behaved) correspond to what \citeauthor{cai2014changes} call
\emph{pointwise differences} \citep[see][section 2.2]{cai2014changes}. As
they point out, you can reconstruct their
function changes from pointwise changes and derivatives, so the two formulations
are equivalent. 

The equivalence of our presentations means that our work should be compatible
with ILC \citep[see][section 3]{cai2014changes}. The derivatives we give in \cref{sec:datalogDifferentiability} are more or
less a ``change semantics'' for Datalog \citep[see][section
3.5]{cai2014changes}. 

\subsubsection{S-acts}
\label{sec:sacts}

S-acts (i.e the category of monoid actions on sets) and their categorical structure have received a fair amount of attention
over the years (\citet{kilp2000monoids} is a good
overview). However, there is a key difference between change actions considered
as a category ($\cat{CAct}$) and the category of
S-acts $\cat{SAct}$: the objects of $\cat{SAct}$ all maintain the same monoid
structure, whereas we are interested in changing both the base set \emph{and} the structure of the action.

There are similarities: if we compare the definition of an ``act-preserving''
homeomorphism in $\cat{SAct}$ \citep[see][]{kilp2000monoids} we can see that the structure is
quite similar to the definition of differentiability:
\begin{displaymath}
  f(a \splus s) = f(a) \splus s
\end{displaymath}
as opposed to
\begin{displaymath}
  f(a \cplus s) = f(a) \cplus \derive{f}(a, s)
\end{displaymath}
That is, we use $\derive{f}$ to transform the action element into the new
monoid, whereas in $\cat{SAct}$ it simply remains the same.

In fact, $\cat{SAct}$ is a subcategory of $\cat{CAct}$, where we only
consider change actions with change set $S$, and the only functions are those
whose derivative is $\lambda a. \lambda d. d$.

\subsubsection{Derivatives of fixpoints}

\citet{arntz2017fixpoints} gives a derivative operator for fixpoints based on
the framework in \citet{cai2014changes}. However, since we have different
notions of function changes, the result is inapplicable as
stated. In addition, we require a somewhat different set of conditions; in particular, we
don't require our changes to always be increasing.

\subsection{Datalog}

\subsubsection{Incremental evaluation}

The earliest interpretation of semi-naive evaluation as a derivative 
appears in \citet{bancilhon1986naive}. The idea of using an approximate derivative
and the requisite soundness condition appears as a throwaway comment in
\citet[][section 3.2.2]{bancilhon1986amateur}, and it would appear that nobody has since
developed that approach.

As far as we know, traditional semi-naive is the state of
the art in incremental, bottom-up, Datalog evaluation, and there are no strategies that
accommodate additional language features such as parity-stratified negation and aggregates.

\subsubsection{Incremental maintenance}

There is existing literature on incremental maintenance of relational algebra
expressions. In particular \citet{griffin1997improved} following
\citet{qian1991incremental} reveal the essential insight that it is necessary to
track both an ``upwards'' and a ``downwards'' difference, and produce a set of
rules that look quite similar to those we derive in \cref{thm:concreteDatalog}.

Where our presentation improves over \citeauthor{griffin1997improved} is mainly in
the genericity of the presentation. Our machinery works for a wider variety of
algebraic structures, and it is clear how the parts of the proof work together
to produce the result. In addition, it is easy to see how to extend the proofs
to cover additional language constructs.

There are some inessential points of difference as well: we work on Datalog,
rather than relational algebra; and we use set semantics rather than bag
semantics. This is largely a matter of convenience: Datalog is an easier
language to work with, and set semantics allows a much wider range of valid
simplifications. However, all the same machinery applies to relational algebra
with bag semantics, it is simply necessary to produce a valid version of
\cref{thm:concreteDatalog}. Since bag semantics also has a complete change
action (using multisets), we can always do this.

We also solve the problem of updating \emph{recursive} expressions. As far as we
know, this is unsolved in general. Most of the attempts to solve it have
focussed on Datalog rather than relational algebra, since Datalog is designed to
make heavy use of recursion.

Several approaches
\citep{gupta1993maintaining,harrison1992maintenance}, most notably DReD
make use of a common tactic: one can get to the new fixed
point by starting from \emph{any} point below it, and then iterating the
semantics again to fixpoint. The approach, then, is to find a way to delete as
few facts as possible to get below the new fixpoint, and then iterate again
(possibly using an incremental version of the semantics).

This is a perfectly reasonable approach, and given a good, domain-specific,
means of getting below the fixpoint, they can be quite efficient.
The main defect of these approaches is that they \emph{are} domain specific,
and hence inflexible with respect to changes in the language or structure,
whereas our approach is quite generic. Although we know of no theoretical reason
why either approach should give superior performance when both are applicable,
an empirical investigation of this could prove interesting.

Other approaches \citep{dong2000incremental,urpi1992method} consider only
restricted subsets of Datalog, or incur other substantial constraints, and our results
are thus significantly more general.

\subsubsection{Embedding Datalog}
\label{sec:embeddingDatalog}

Datafun (\citet{arntz2016datafun}) is a functional programming language that embeds
Datalog, allowing significant improvements in genericity, such as the use of
higher-order functions. Since we have directly defined a change action and
derivative operator for Datalog, our work could be used as a ``plugin'' in the sense
of \citeauthor{cai2014changes}, allowing Datafun to compute its internal fixpoints
incrementally, but also allowing Datafun expressions to be fully incrementally maintained.

In a different direction, \citet{cathcartburn2018hochc} have proposed
\emph{higher-order constrained Horn clauses}
(HoCHC), a new class of constraints for the automatic verification of higher-order programs.
HoCHC may be viewed as a higher-order extension of Datalog. 
Change actions can be readily applied to organise an efficient semi-naive method for solving HoCHC systems.

\subsection{Differential $\lambda$-calculus}

Another setting where derivatives of arbitrary higher-order programs have been studied
is the \emph{differential $\lambda$-calculus} \citep{ehrhard2003differential,ehrhard2017introduction}.
This is a higher-order, simply-typed $\lambda$-calculus which allows for computing the derivative of a function, in a similar
way to the notion of derivative in Cai's work and the present paper.

While there are clear similarities between the two systems, 
the most important difference is the properties of the derivatives themselves:
in the differential $\lambda$-calculus, derivatives are guaranteed to be linear
in their second argument, whereas in our approach derivatives do not have this restriction 
but are instead required to satisfy a strong relation to the function
that is being differentiated (see \cref{def:derivative}).

Families of denotational models for the differential $\lambda$-calculus have been
studied in depth
\citep{bucciarelli2010categorical,blute2010convenient,cockett2016categorical,kerjean2016mackey},
and the relationship between these and change actions is the subject of ongoing work.

\subsection{Higher-order automatic differentiation}

Automatic differentiation \citep{griewank2008evaluating} is a technique that allows
for efficiently computing the derivative of arbitrary programs, with
applications in probabilistic modeling \citep{kucukelbir2017automatic}
and machine learning \citep{baydin2014automatic} among other areas. In recent times, this technique has been successfully
applied to higher-order languages \citep{siskind2008nesting,baydin2016diffsharp}.
While some approaches have been suggested \citep{manzyuk2012simply,kelly2016evolving}, a general
theoretical framework for this technique is still a matter of open research. 

To this purpose, some authors have proposed the incremental $\lambda$-calculus
as a foundational framework on which models of automatic differentiation can
be based \citep{kelly2016evolving}. We believe our change actions are better suited
to this purpose than the incremental $\lambda$-calculus, since one can easily give them a
synthetic differential geometric reading (by interpreting $\cstr{A}$ as an Euclidean module and $\changes{A}$
as its corresponding spectrum, for example).

\section{Future work}

Our work opens a number of avenues for future investigation.

\subsection{The category of change actions}

First, there is a category $\cat{CAct}$ of change actions and differentiable
functions between them, since the identity function is differentiable and the chain
rule guarantees that composition preserves differentiability. The product and sum change
actions described in section \cref{sec:prodsum} correspond to products and
coproducts in this category. Furthermore, \cref{thm:functionalCAct} can be
read as showing that complete change actions form an exponential ideal in $\cat{CAct}$.

A simple result shows that the category $\cat{CAct}$ is remarkably well-behaved:

\begin{theorem}
  The category $\cat{CAct}$ is equivalent to the category $\cat{PreOrd}$ of
  preorders and monotone functions.
\end{theorem}

This has the consequence that $\cat{CAct}$ is Cartesian closed.
This result, however, isn't very helpful in practice
as the resulting exponentials in $\cat{CAct}$ are not computable. A more thorough study
of the category $\cat{CAct}$ could lead to a presentation of exponentials that is more
suitable to actual computation.

Fortunately, under some circumstances exponential objects in $\cat{CAct}$ correspond
to the pointwise functional change actions we defined earlier (in \cref{sec:pointwiseFunctional}).

\begin{defn}
  \label{def:convenientChangeActions}
  A change action $\cstr{B}$ is \emph{convenient} if every exponential object
  $\exponential{\cstr{A}}{\cstr{B}}$ is isomorphic to a pointwise functional change action.
\end{defn}

Fortunately, convenient change actions form a relatively large subcategory of
$\cat{CAct}$. In particular, complete change actions are convenient.

Since the product and exponential of complete change actions are also complete,  the complete change
actions are a Cartesian closed full subcategory of $\cat{CAct}$ which is, in fact equivalent
to $\cat{Set}$ (as every set-theoretic function between complete change actions is differentiable).

Furthermore, it can be shown that the product and exponential of two convenient
change actions are in turn convenient change actions, and hence convenient
change actions are themselves a Cartesian closed subcategory of $\cat{CAct}$. We
believe these can therefore provide a suitable model for general higher-order
incremental computation in the future.

These preliminary results all suggest that the category $\cat{CAct}$ is an
interesting subject of study in itself. Many important questions, especially
with respect to the structure of exponentials and the existence of limits and
colimits remain unsettled.

\subsection{Change actions as 2-categories}

There is an interesting connection between the category $\cat{CAct}$ and the 2-category 
$\cat{Cat}$ of categories and functors: given a change action $\cstr{A}$ one can consider
the category that has elements of $A$ as objects and a morphism $f_{\change a} : a \ra b$ for
every change $\change a \in \changes A$ such that $a \oplus \change a = b$. If one requires
that derivatives verify the additional conditions in the footnote to \cref{def:derivative}, then functions that
admit derivatives correspond exactly to functors. This naturally leads to consider dependently-typed
generalizations of change actions where different elements $a, b \in A$ have different change
sets $\changes_aA, \changes_aB$, much like in \citeauthor{cai2014changes}'s original presentation.

\subsection{Change actions on arbitrary categories}

A more abstract approach is also possible: throughout this work we have restricted ourselves
to change actions defined over the category of sets, but most of the definitions and results
presented here could be easily generalized to any category with products. The resulting theory
is the subject of ongoing work by the authors.

Of those, the theory of change actions on the category of domains is of
particular interest to us. Since domains are used to model programming language
semantics in the presence of general recursion, a theory of change actions on
the category of domains could open up opportunities for incremental evaluation
of many programming languages, even those that do not fit into the model of
\citeauthor{cai2014changes}'s ILC. Our fixpoint theorems are proven over dcpos
in general, and we believe these to be a crucial first step into a more general
theory of a hypothetical incremental extension to PCF.

Additionally, we have only begun to explore the tantalizing connection between
change actions, the ILC and synthetic differential geometry, and a denotational
semantics for the ILC based on smooth spaces is the subject of ongoing research.
We believe many concepts from standard differential geometry, like gradients,
vector fields, curves and flows can be defined in general on change actions, which
could lead to developments in differentiable programming languages.

Finally, combining our concrete Datalog derivatives with a system similar to ILC
in a language such as Datafun would be an exciting demonstration of the compositional
power of this approach.

\section{Conclusions}

We have presented change actions and their properties, and used them to provide
novel, compositional, strategies for incrementally evaluating and maintaining recursive functions, in
particular the semantics of Datalog.


\begin{acks}

We would like to thank Semmle Ltd. for supporting this research, as well as Pavel
Avgustinov, Aditya Sharad, Max Sch\"afer, Katriel Cohn-Gordon, Anders
Schack-Mulligen, and Simon Peyton Jones for their
helpful comments on the manuscript.

\end{acks}

\bibliographystyle{ACM-Reference-Format}
\bibliography{paper}


\begin{thebibliography}{41}


\ifx \showCODEN    \undefined \def \showCODEN     #1{\unskip}     \fi
\ifx \showDOI      \undefined \def \showDOI       #1{#1}\fi
\ifx \showISBNx    \undefined \def \showISBNx     #1{\unskip}     \fi
\ifx \showISBNxiii \undefined \def \showISBNxiii  #1{\unskip}     \fi
\ifx \showISSN     \undefined \def \showISSN      #1{\unskip}     \fi
\ifx \showLCCN     \undefined \def \showLCCN      #1{\unskip}     \fi
\ifx \shownote     \undefined \def \shownote      #1{#1}          \fi
\ifx \showarticletitle \undefined \def \showarticletitle #1{#1}   \fi
\ifx \showURL      \undefined \def \showURL       {\relax}        \fi
\providecommand\bibfield[2]{#2}
\providecommand\bibinfo[2]{#2}
\providecommand\natexlab[1]{#1}
\providecommand\showeprint[2][]{arXiv:#2}

\bibitem[\protect\citeauthoryear{Abiteboul, Hull, and Vianu}{Abiteboul
  et~al\mbox{.}}{1995}]%
        {abiteboul1995foundations}
\bibfield{author}{\bibinfo{person}{Serge Abiteboul}, \bibinfo{person}{Richard
  Hull}, {and} \bibinfo{person}{Victor Vianu}.}
  \bibinfo{year}{1995}\natexlab{}.
\newblock \bibinfo{booktitle}{\emph{Foundations of databases: the logical
  level}}.
\newblock \bibinfo{publisher}{Addison-Wesley Longman Publishing Co., Inc.}
\newblock


\bibitem[\protect\citeauthoryear{Abramsky and Jung}{Abramsky and Jung}{1994}]%
        {abramsky1994domain}
\bibfield{author}{\bibinfo{person}{Samson Abramsky} {and}
  \bibinfo{person}{Achim Jung}.} \bibinfo{year}{1994}\natexlab{}.
\newblock \showarticletitle{Domain theory}. In
  \bibinfo{booktitle}{\emph{Handbook of logic in computer science}}. Oxford
  University Press.
\newblock


\bibitem[\protect\citeauthoryear{Arntzenius}{Arntzenius}{2017}]%
        {arntz2017fixpoints}
\bibfield{author}{\bibinfo{person}{Michael Arntzenius}.}
  \bibinfo{year}{2017}\natexlab{}.
\newblock \bibinfo{title}{Static differentiation of monotone fixpoints}.
  (\bibinfo{year}{2017}).
\newblock
\urldef\tempurl%
\url{http://www.rntz.net/files/fixderiv.pdf}
\showURL{%
\tempurl}


\bibitem[\protect\citeauthoryear{Arntzenius and Krishnaswami}{Arntzenius and
  Krishnaswami}{2016}]%
        {arntz2016datafun}
\bibfield{author}{\bibinfo{person}{Michael Arntzenius} {and}
  \bibinfo{person}{Neelakantan~R Krishnaswami}.}
  \bibinfo{year}{2016}\natexlab{}.
\newblock \showarticletitle{Datafun: a functional Datalog}. In
  \bibinfo{booktitle}{\emph{Proceedings of the 21st ACM SIGPLAN International
  Conference on Functional Programming}}. ACM, \bibinfo{pages}{214--227}.
\newblock


\bibitem[\protect\citeauthoryear{Avgustinov, de~Moor, Jones, and
  Sch{\"a}fer}{Avgustinov et~al\mbox{.}}{2016}]%
        {avgustinov2016ql}
\bibfield{author}{\bibinfo{person}{Pavel Avgustinov}, \bibinfo{person}{Oege de
  Moor}, \bibinfo{person}{Michael~Peyton Jones}, {and} \bibinfo{person}{Max
  Sch{\"a}fer}.} \bibinfo{year}{2016}\natexlab{}.
\newblock \showarticletitle{QL: Object-oriented Queries on Relational Data}. In
  \bibinfo{booktitle}{\emph{LIPIcs-Leibniz International Proceedings in
  Informatics}}, Vol.~\bibinfo{volume}{56}. Schloss Dagstuhl-Leibniz-Zentrum
  fuer Informatik.
\newblock


\bibitem[\protect\citeauthoryear{Bancilhon}{Bancilhon}{1986}]%
        {bancilhon1986naive}
\bibfield{author}{\bibinfo{person}{Francois Bancilhon}.}
  \bibinfo{year}{1986}\natexlab{}.
\newblock \showarticletitle{Naive evaluation of recursively defined relations}.
\newblock In \bibinfo{booktitle}{\emph{On Knowledge Base Management Systems}}.
  \bibinfo{publisher}{Springer}, \bibinfo{pages}{165--178}.
\newblock


\bibitem[\protect\citeauthoryear{Bancilhon and Ramakrishnan}{Bancilhon and
  Ramakrishnan}{1986}]%
        {bancilhon1986amateur}
\bibfield{author}{\bibinfo{person}{Francois Bancilhon} {and}
  \bibinfo{person}{Raghu Ramakrishnan}.} \bibinfo{year}{1986}\natexlab{}.
\newblock \bibinfo{booktitle}{\emph{An amateur's introduction to recursive
  query processing strategies}}. Vol.~\bibinfo{volume}{15}.
\newblock \bibinfo{publisher}{ACM}.
\newblock


\bibitem[\protect\citeauthoryear{Baydin and Pearlmutter}{Baydin and
  Pearlmutter}{2014}]%
        {baydin2014automatic}
\bibfield{author}{\bibinfo{person}{Atilim~Gunes Baydin} {and}
  \bibinfo{person}{Barak~A Pearlmutter}.} \bibinfo{year}{2014}\natexlab{}.
\newblock \showarticletitle{Automatic differentiation of algorithms for machine
  learning}.
\newblock \bibinfo{journal}{\emph{arXiv preprint arXiv:1404.7456}}
  (\bibinfo{year}{2014}).
\newblock


\bibitem[\protect\citeauthoryear{Baydin, Pearlmutter, and Siskind}{Baydin
  et~al\mbox{.}}{2016}]%
        {baydin2016diffsharp}
\bibfield{author}{\bibinfo{person}{At{\i}l{\i}m~G{\"u}ne{\c{s}} Baydin},
  \bibinfo{person}{Barak~A Pearlmutter}, {and} \bibinfo{person}{Jeffrey~Mark
  Siskind}.} \bibinfo{year}{2016}\natexlab{}.
\newblock \showarticletitle{DiffSharp: An AD Library for. NET Languages}.
\newblock \bibinfo{journal}{\emph{arXiv preprint arXiv:1611.03423}}
  (\bibinfo{year}{2016}).
\newblock


\bibitem[\protect\citeauthoryear{Blute, Ehrhard, and Tasson}{Blute
  et~al\mbox{.}}{2010}]%
        {blute2010convenient}
\bibfield{author}{\bibinfo{person}{Richard Blute}, \bibinfo{person}{Thomas
  Ehrhard}, {and} \bibinfo{person}{Christine Tasson}.}
  \bibinfo{year}{2010}\natexlab{}.
\newblock \showarticletitle{A convenient differential category}.
\newblock \bibinfo{journal}{\emph{arXiv preprint arXiv:1006.3140}}
  (\bibinfo{year}{2010}).
\newblock


\bibitem[\protect\citeauthoryear{Bucciarelli, Ehrhard, and
  Manzonetto}{Bucciarelli et~al\mbox{.}}{2010}]%
        {bucciarelli2010categorical}
\bibfield{author}{\bibinfo{person}{Antonio Bucciarelli},
  \bibinfo{person}{Thomas Ehrhard}, {and} \bibinfo{person}{Giulio Manzonetto}.}
  \bibinfo{year}{2010}\natexlab{}.
\newblock \showarticletitle{Categorical models for simply typed resource
  calculi}.
\newblock \bibinfo{journal}{\emph{Electronic Notes in Theoretical Computer
  Science}}  \bibinfo{volume}{265} (\bibinfo{year}{2010}),
  \bibinfo{pages}{213--230}.
\newblock


\bibitem[\protect\citeauthoryear{Cai, Giarrusso, Rendel, and Ostermann}{Cai
  et~al\mbox{.}}{2014}]%
        {cai2014changes}
\bibfield{author}{\bibinfo{person}{Yufei Cai}, \bibinfo{person}{Paolo~G
  Giarrusso}, \bibinfo{person}{Tillmann Rendel}, {and} \bibinfo{person}{Klaus
  Ostermann}.} \bibinfo{year}{2014}\natexlab{}.
\newblock \showarticletitle{A theory of changes for higher-order languages:
  Incrementalizing $\lambda$-calculi by static differentiation}. In
  \bibinfo{booktitle}{\emph{ACM SIGPLAN Notices}}, Vol.~\bibinfo{volume}{49}.
  ACM, \bibinfo{pages}{145--155}.
\newblock


\bibitem[\protect\citeauthoryear{{Cathcart Burn}, Ong, and Ramsay}{{Cathcart
  Burn} et~al\mbox{.}}{2018}]%
        {cathcartburn2018hochc}
\bibfield{author}{\bibinfo{person}{Toby {Cathcart Burn}},
  \bibinfo{person}{C.{-}H.~Luke Ong}, {and} \bibinfo{person}{Steven~J.
  Ramsay}.} \bibinfo{year}{2018}\natexlab{}.
\newblock \showarticletitle{Higher-order constrained horn clauses for
  verification}.
\newblock \bibinfo{journal}{\emph{{PACMPL}}} \bibinfo{volume}{2},
  \bibinfo{number}{{POPL}} (\bibinfo{year}{2018}),
  \bibinfo{pages}{11:1--11:28}.
\newblock
\urldef\tempurl%
\url{https://doi.org/10.1145/3158099}
\showDOI{\tempurl}


\bibitem[\protect\citeauthoryear{Cockett and Gallagher}{Cockett and
  Gallagher}{2016}]%
        {cockett2016categorical}
\bibfield{author}{\bibinfo{person}{J~Robin~B Cockett} {and} \bibinfo{person}{JD
  Gallagher}.} \bibinfo{year}{2016}\natexlab{}.
\newblock \showarticletitle{Categorical models of the differential
  $\lambda$-calculus revisited}.
\newblock \bibinfo{journal}{\emph{Electronic Notes in Theoretical Computer
  Science}}  \bibinfo{volume}{325} (\bibinfo{year}{2016}),
  \bibinfo{pages}{63--83}.
\newblock


\bibitem[\protect\citeauthoryear{Compton}{Compton}{1994}]%
        {compton1994stratified}
\bibfield{author}{\bibinfo{person}{Kevin~J Compton}.}
  \bibinfo{year}{1994}\natexlab{}.
\newblock \showarticletitle{Stratified least fixpoint logic}.
\newblock \bibinfo{journal}{\emph{Theoretical Computer Science}}
  \bibinfo{volume}{131}, \bibinfo{number}{1} (\bibinfo{year}{1994}),
  \bibinfo{pages}{95--120}.
\newblock


\bibitem[\protect\citeauthoryear{Datomic}{Datomic}{[n. d.]}]%
        {datomicWebsite}
Datomic \bibinfo{year}{[n. d.]}\natexlab{}.
\newblock \bibinfo{title}{Datomic website}.
\newblock
\newblock
\urldef\tempurl%
\url{https://www.datomic.com}
\showURL{%
\tempurl}
\newblock
\shownote{Accessed: 2018-01-01.}


\bibitem[\protect\citeauthoryear{de~Moor and Baars}{de~Moor and Baars}{2013}]%
        {demoor2013aggregates}
\bibfield{author}{\bibinfo{person}{Oege de Moor} {and} \bibinfo{person}{Arthur
  Baars}.} \bibinfo{year}{2013}\natexlab{}.
\newblock \showarticletitle{Doing a Doaitse: Simple Recursive Aggregates in
  Datalog}.
\newblock In \bibinfo{booktitle}{\emph{Liber Amicorum for Doaitse Swierstra}}.
  \bibinfo{pages}{207--216}.
\newblock
\urldef\tempurl%
\url{http://www.staff.science.uu.nl/~hage0101/liberdoaitseswierstra.pdf}
\showURL{%
\tempurl}
\newblock
\shownote{Accessed: 2018-01-01.}


\bibitem[\protect\citeauthoryear{Dong and Su}{Dong and Su}{2000}]%
        {dong2000incremental}
\bibfield{author}{\bibinfo{person}{Guozhu Dong} {and} \bibinfo{person}{Jianwen
  Su}.} \bibinfo{year}{2000}\natexlab{}.
\newblock \showarticletitle{Incremental maintenance of recursive views using
  relational calculus/SQL}.
\newblock \bibinfo{journal}{\emph{ACM SIGMOD Record}} \bibinfo{volume}{29},
  \bibinfo{number}{1} (\bibinfo{year}{2000}), \bibinfo{pages}{44--51}.
\newblock


\bibitem[\protect\citeauthoryear{Ehrhard}{Ehrhard}{2017}]%
        {ehrhard2017introduction}
\bibfield{author}{\bibinfo{person}{Thomas Ehrhard}.}
  \bibinfo{year}{2017}\natexlab{}.
\newblock \showarticletitle{An introduction to Differential Linear Logic:
  proof-nets, models and antiderivatives}.
\newblock \bibinfo{journal}{\emph{Mathematical Structures in Computer Science}}
  (\bibinfo{year}{2017}), \bibinfo{pages}{1--66}.
\newblock


\bibitem[\protect\citeauthoryear{Ehrhard and Regnier}{Ehrhard and
  Regnier}{2003}]%
        {ehrhard2003differential}
\bibfield{author}{\bibinfo{person}{Thomas Ehrhard} {and}
  \bibinfo{person}{Laurent Regnier}.} \bibinfo{year}{2003}\natexlab{}.
\newblock \showarticletitle{The differential lambda-calculus}.
\newblock \bibinfo{journal}{\emph{Theoretical Computer Science}}
  \bibinfo{volume}{309}, \bibinfo{number}{1-3} (\bibinfo{year}{2003}),
  \bibinfo{pages}{1--41}.
\newblock


\bibitem[\protect\citeauthoryear{Griewank and Walther}{Griewank and
  Walther}{2008}]%
        {griewank2008evaluating}
\bibfield{author}{\bibinfo{person}{Andreas Griewank} {and}
  \bibinfo{person}{Andrea Walther}.} \bibinfo{year}{2008}\natexlab{}.
\newblock \bibinfo{booktitle}{\emph{Evaluating derivatives: principles and
  techniques of algorithmic differentiation}}. Vol.~\bibinfo{volume}{105}.
\newblock \bibinfo{publisher}{Siam}.
\newblock


\bibitem[\protect\citeauthoryear{Griffin, Libkin, and Trickey}{Griffin
  et~al\mbox{.}}{1997}]%
        {griffin1997improved}
\bibfield{author}{\bibinfo{person}{Timothy Griffin}, \bibinfo{person}{Leonid
  Libkin}, {and} \bibinfo{person}{Howard Trickey}.}
  \bibinfo{year}{1997}\natexlab{}.
\newblock \showarticletitle{An improved algorithm for the incremental
  recomputation of active relational expressions}.
\newblock \bibinfo{journal}{\emph{IEEE Transactions on Knowledge \& Data
  Engineering}} \bibinfo{number}{3} (\bibinfo{year}{1997}),
  \bibinfo{pages}{508--511}.
\newblock


\bibitem[\protect\citeauthoryear{Gupta, Mumick, and Subrahmanian}{Gupta
  et~al\mbox{.}}{1993}]%
        {gupta1993maintaining}
\bibfield{author}{\bibinfo{person}{Ashish Gupta},
  \bibinfo{person}{Inderpal~Singh Mumick}, {and}
  \bibinfo{person}{Venkatramanan~Siva Subrahmanian}.}
  \bibinfo{year}{1993}\natexlab{}.
\newblock \showarticletitle{Maintaining views incrementally}.
\newblock \bibinfo{journal}{\emph{ACM SIGMOD Record}} \bibinfo{volume}{22},
  \bibinfo{number}{2} (\bibinfo{year}{1993}), \bibinfo{pages}{157--166}.
\newblock


\bibitem[\protect\citeauthoryear{Halpin and Rugaber}{Halpin and
  Rugaber}{2014}]%
        {halpin2014logiql}
\bibfield{author}{\bibinfo{person}{Terry Halpin} {and} \bibinfo{person}{Spencer
  Rugaber}.} \bibinfo{year}{2014}\natexlab{}.
\newblock \bibinfo{booktitle}{\emph{LogiQL: A Query Language for Smart
  Databases}}.
\newblock \bibinfo{publisher}{CRC Press}.
\newblock


\bibitem[\protect\citeauthoryear{Harrison and Dietrich}{Harrison and
  Dietrich}{1992}]%
        {harrison1992maintenance}
\bibfield{author}{\bibinfo{person}{John~V Harrison} {and}
  \bibinfo{person}{Suzanne~W Dietrich}.} \bibinfo{year}{1992}\natexlab{}.
\newblock \showarticletitle{Maintenance of Materialized Views in a Deductive
  Database: An Update Propagation Approach.}. In
  \bibinfo{booktitle}{\emph{Workshop on Deductive Databases, JICSLP}}.
  \bibinfo{pages}{56--65}.
\newblock


\bibitem[\protect\citeauthoryear{Kelly, Pearlmutter, and Siskind}{Kelly
  et~al\mbox{.}}{2016}]%
        {kelly2016evolving}
\bibfield{author}{\bibinfo{person}{Robert Kelly}, \bibinfo{person}{Barak~A
  Pearlmutter}, {and} \bibinfo{person}{Jeffrey~Mark Siskind}.}
  \bibinfo{year}{2016}\natexlab{}.
\newblock \showarticletitle{Evolving the Incremental $\lambda$ Calculus into a
  Model of Forward Automatic Differentiation (AD)}.
\newblock \bibinfo{journal}{\emph{arXiv preprint arXiv:1611.03429}}
  (\bibinfo{year}{2016}).
\newblock


\bibitem[\protect\citeauthoryear{Kerjean and Tasson}{Kerjean and
  Tasson}{2016}]%
        {kerjean2016mackey}
\bibfield{author}{\bibinfo{person}{Marie Kerjean} {and}
  \bibinfo{person}{Christine Tasson}.} \bibinfo{year}{2016}\natexlab{}.
\newblock \showarticletitle{Mackey-complete spaces and power series--A
  topological model of Differential Linear Logic}.
\newblock \bibinfo{journal}{\emph{Mathematical Structures in Computer Science}}
  (\bibinfo{year}{2016}), \bibinfo{pages}{1--36}.
\newblock


\bibitem[\protect\citeauthoryear{Kilp, Knauer, and Mikhalev}{Kilp
  et~al\mbox{.}}{2000}]%
        {kilp2000monoids}
\bibfield{author}{\bibinfo{person}{Mati Kilp}, \bibinfo{person}{Ulrich Knauer},
  {and} \bibinfo{person}{Alexander~V Mikhalev}.}
  \bibinfo{year}{2000}\natexlab{}.
\newblock \bibinfo{booktitle}{\emph{Monoids, Acts and Categories: With
  Applications to Wreath Products and Graphs. A Handbook for Students and
  Researchers}}. Vol.~\bibinfo{volume}{29}.
\newblock \bibinfo{publisher}{Walter de Gruyter}.
\newblock


\bibitem[\protect\citeauthoryear{Kucukelbir, Tran, Ranganath, Gelman, and
  Blei}{Kucukelbir et~al\mbox{.}}{2017}]%
        {kucukelbir2017automatic}
\bibfield{author}{\bibinfo{person}{Alp Kucukelbir}, \bibinfo{person}{Dustin
  Tran}, \bibinfo{person}{Rajesh Ranganath}, \bibinfo{person}{Andrew Gelman},
  {and} \bibinfo{person}{David~M Blei}.} \bibinfo{year}{2017}\natexlab{}.
\newblock \showarticletitle{Automatic differentiation variational inference}.
\newblock \bibinfo{journal}{\emph{The Journal of Machine Learning Research}}
  \bibinfo{volume}{18}, \bibinfo{number}{1} (\bibinfo{year}{2017}),
  \bibinfo{pages}{430--474}.
\newblock


\bibitem[\protect\citeauthoryear{Logicblox Inc.}{Logicblox Inc.}{[n. d.]}]%
        {logicbloxWebsite}
Logicblox Inc. \bibinfo{year}{[n. d.]}\natexlab{}.
\newblock \bibinfo{title}{Logicblox Inc. website}.
\newblock
\newblock
\urldef\tempurl%
\url{http://www.logicblox.com}
\showURL{%
\tempurl}
\newblock
\shownote{Accessed: 2018-01-01.}


\bibitem[\protect\citeauthoryear{Manzyuk}{Manzyuk}{2012}]%
        {manzyuk2012simply}
\bibfield{author}{\bibinfo{person}{Oleksandr Manzyuk}.}
  \bibinfo{year}{2012}\natexlab{}.
\newblock \showarticletitle{A simply typed $\lambda$-calculus of forward
  automatic differentiation}.
\newblock \bibinfo{journal}{\emph{Electronic Notes in Theoretical Computer
  Science}}  \bibinfo{volume}{286} (\bibinfo{year}{2012}),
  \bibinfo{pages}{257--272}.
\newblock


\bibitem[\protect\citeauthoryear{Qian and Wiederhold}{Qian and
  Wiederhold}{1991}]%
        {qian1991incremental}
\bibfield{author}{\bibinfo{person}{Xiaolei Qian} {and} \bibinfo{person}{Gio
  Wiederhold}.} \bibinfo{year}{1991}\natexlab{}.
\newblock \showarticletitle{Incremental recomputation of active relational
  expressions}.
\newblock \bibinfo{journal}{\emph{Knowledge and Data Engineering, IEEE
  Transactions on}} \bibinfo{volume}{3}, \bibinfo{number}{3}
  (\bibinfo{year}{1991}), \bibinfo{pages}{337--341}.
\newblock


\bibitem[\protect\citeauthoryear{S{\'a}enz-P{\'e}rez}{S{\'a}enz-P{\'e}rez}{2011}]%
        {saenz2011deductive}
\bibfield{author}{\bibinfo{person}{Fernando S{\'a}enz-P{\'e}rez}.}
  \bibinfo{year}{2011}\natexlab{}.
\newblock \showarticletitle{DES: A deductive database system}.
\newblock \bibinfo{journal}{\emph{Electronic notes in theoretical computer
  science}}  \bibinfo{volume}{271} (\bibinfo{year}{2011}),
  \bibinfo{pages}{63--78}.
\newblock


\bibitem[\protect\citeauthoryear{Sch{\"a}fer and de~Moor}{Sch{\"a}fer and
  de~Moor}{2010}]%
        {schafer2010type}
\bibfield{author}{\bibinfo{person}{Max Sch{\"a}fer} {and} \bibinfo{person}{Oege
  de Moor}.} \bibinfo{year}{2010}\natexlab{}.
\newblock \showarticletitle{Type inference for datalog with complex type
  hierarchies}. In \bibinfo{booktitle}{\emph{ACM Sigplan Notices}},
  Vol.~\bibinfo{volume}{45}. ACM, \bibinfo{pages}{145--156}.
\newblock


\bibitem[\protect\citeauthoryear{Scholz, Jordan, Suboti{\'c}, and
  Westmann}{Scholz et~al\mbox{.}}{2016}]%
        {scholz2016fast}
\bibfield{author}{\bibinfo{person}{Bernhard Scholz}, \bibinfo{person}{Herbert
  Jordan}, \bibinfo{person}{Pavle Suboti{\'c}}, {and} \bibinfo{person}{Till
  Westmann}.} \bibinfo{year}{2016}\natexlab{}.
\newblock \showarticletitle{On fast large-scale program analysis in datalog}.
  In \bibinfo{booktitle}{\emph{Proceedings of the 25th International Conference
  on Compiler Construction}}. ACM, \bibinfo{pages}{196--206}.
\newblock


\bibitem[\protect\citeauthoryear{Semmle Ltd.}{Semmle Ltd.}{[n. d.]}]%
        {semmleWebsite}
Semmle Ltd. \bibinfo{year}{[n. d.]}\natexlab{}.
\newblock \bibinfo{title}{Semmle Ltd. website}.
\newblock
\newblock
\urldef\tempurl%
\url{https://semmle.com}
\showURL{%
\tempurl}
\newblock
\shownote{Accessed: 2018-01-01.}


\bibitem[\protect\citeauthoryear{Sereni, Avgustinov, and De~Moor}{Sereni
  et~al\mbox{.}}{2008}]%
        {sereni2008adding}
\bibfield{author}{\bibinfo{person}{Damien Sereni}, \bibinfo{person}{Pavel
  Avgustinov}, {and} \bibinfo{person}{Oege De~Moor}.}
  \bibinfo{year}{2008}\natexlab{}.
\newblock \showarticletitle{Adding magic to an optimising datalog compiler}. In
  \bibinfo{booktitle}{\emph{Proceedings of the 2008 ACM SIGMOD international
  conference on Management of data}}. ACM, \bibinfo{pages}{553--566}.
\newblock


\bibitem[\protect\citeauthoryear{Siskind and Pearlmutter}{Siskind and
  Pearlmutter}{2008}]%
        {siskind2008nesting}
\bibfield{author}{\bibinfo{person}{Jeffrey~Mark Siskind} {and}
  \bibinfo{person}{Barak~A Pearlmutter}.} \bibinfo{year}{2008}\natexlab{}.
\newblock \showarticletitle{Nesting forward-mode AD in a functional framework}.
\newblock \bibinfo{journal}{\emph{Higher-Order and Symbolic Computation}}
  \bibinfo{volume}{21}, \bibinfo{number}{4} (\bibinfo{year}{2008}),
  \bibinfo{pages}{361--376}.
\newblock


\bibitem[\protect\citeauthoryear{Souffle}{Souffle}{[n. d.]}]%
        {souffleWebsite}
Souffle \bibinfo{year}{[n. d.]}\natexlab{}.
\newblock \bibinfo{title}{Souffle language website}.
\newblock
\newblock
\urldef\tempurl%
\url{http://souffle-lang.org}
\showURL{%
\tempurl}
\newblock
\shownote{Accessed: 2018-01-01.}


\bibitem[\protect\citeauthoryear{Urpi and Olive}{Urpi and Olive}{1992}]%
        {urpi1992method}
\bibfield{author}{\bibinfo{person}{Toni Urpi} {and} \bibinfo{person}{Antoni
  Olive}.} \bibinfo{year}{1992}\natexlab{}.
\newblock \showarticletitle{A method for change computation in deductive
  databases}. In \bibinfo{booktitle}{\emph{VLDB}}, Vol.~\bibinfo{volume}{92}.
  \bibinfo{pages}{225--237}.
\newblock


\bibitem[\protect\citeauthoryear{Winskel}{Winskel}{1993}]%
        {winskel1993formal}
\bibfield{author}{\bibinfo{person}{Glynn Winskel}.}
  \bibinfo{year}{1993}\natexlab{}.
\newblock \bibinfo{booktitle}{\emph{The formal semantics of programming
  languages: an introduction}}.
\newblock \bibinfo{publisher}{MIT press}.
\newblock


\end{thebibliography}

\clearpage
\appendix
\appendixpage
\section{Proofs}

\subsection{Change actions and derivatives}

\products*
\begin{proof}
  \label{prf:products}
  Let $\cstr{Y}$ be a change action, and $f_1: \cstr{Y} \rightarrow \cstr{A}$, $f_2: \cstr{Y}
  \rightarrow \cstr{B}$ be morphisms.

  Then the product morphism in $\cat{Set}$, $\pair{f_1}{f_2}$ is the product
  morphism in $\cat{CAct}$. It can easily be
  shown that $\pair{\derive{f_1}}{\derive{f_2}}$ is a derivative of $\pair{f_1}{f_2}$,
  hence $\pair{f_1}{f_2}$ is a morphism in $\cat{SAct}$.

  Commutativity and uniqueness follow from the corresponding properties of the
  product in the $\cat{Set}$.
\end{proof}

\disjointUnions*
\begin{proof}
  \label{prf:disjointUnions}
  Let $\cstr{Y}$ be a change action, and $f_1 : \cstr{A} \rightarrow \cstr{Y}$, $f_2 : \cstr{B}
  \rightarrow \cstr{Y}$ be differentiable functions.

  As before, it suffices to prove that the universal function $[f_1, f_2]$ in $\cat{Set}$ is a differentiable
  function from $\cstruct{A + B}{\changes{A} \times \changes{B}}{\cplus_{A + B}}$ into $Y$. It's easy to see
  that the following morphism is a derivative:
  \begin{align*}
    \derive{[f_1, f_2]} (i_1 a, (\change{a}, \change{b})) &\defeq f_1'(a, \change{a})\\
    \derive{[f_1, f_2]} (i_2 b, (\change{a}, \change{b})) &\defeq f_2'(b, \change{b})
  \end{align*}
\end{proof}

\subsection{Posets and Boolean algebras}

\lsuperpose*
\begin{proof}
  \label{prf:lsuperpose}
  We show that the monoid action property holds:
  \begin{align*}
    &a \twist \left[(p, q) \splus (r, s)\right]\\
    &= a \twist ((p \wedge \neg s) \vee r, (q \wedge \neg r) \vee s)\\
    &= \left(
      a \vee
      \left(
        \left(
          p \wedge \neg s
        \right)
        \vee r
      \right)
    \right)
    \wedge \neg
    \left(
      \left(
        q \wedge \neg r
      \right)
      \vee s
    \right)\\
    &= \left(
      \left(
        \left(
          a \vee p
        \right)
        \wedge
        \left(
          a \vee \neg s
        \right)
      \right)
      \vee r
    \right)
    \wedge
    \left(
      \neg q \vee r
    \right)
    \wedge
    \neg s
    \tag{distributing $\vee$ over $\wedge$, applying de Morgan rules}\\
    &= \left(
      \left(
        \left(
          a \vee p
        \right)
        \wedge
        \left(
          a \vee \neg s
        \right)
        \wedge
        \neg q
      \right)
      \vee r
    \right)
    \wedge
    \neg s
    \tag{un-distributing $\vee$ over $\wedge$ }\\
    &= \left(
      \left(
        \left(
          a \vee p
        \right)
        \wedge
        \neg q
      \right)
      \vee r
    \right)
    \wedge
    \neg s
    \tag{$r \rightarrow \neg s$}\\
    &= a \twist (p, q) \twist (r, s)
  \end{align*}

  It is easy to show that $\twist$ is well-defined, by showing that $(p,q)
  \twist (r, s) \in L \disjointTimes L$ if $(p, q), (r,s) \in L \disjointTimes L$.

  Completeness is easy to show.
\end{proof}

\booleanAlgebraKernels*
\begin{proof}
  \label{prf:booleanAlgebraKernels}
  
  First we show that for any $p, q, r, s \in L$,
  $p \leq r$ and $q \leq s$ implies $(p, q) \kernelOrder (r, s)$. We
  note that the condition implies
  \begin{align*}
    &(p, q) \splus_\superpose (r,s)\\
    &= ((p \wedge \neg s) \vee r, (q \wedge \neg r) \vee s)\\
    &= (r,s)
  \end{align*}
  Let $a, b \in L$. Then
  \begin{align*}
    &a \twist (p, q) = b \twist (p, q)\\
    &\Rightarrow a \twist (p, q) \twist (r, s) = b \twist (p, q) \splus_\superpose (r, s)\\
    &\Rightarrow a \twist (r, s) = b \twist (r, s)
  \end{align*}
  Hence $\kernel_{(p,q)} \subseteq \kernel_{(r,s)}$, so $(p, q) \kernelOrder (r, s)$. The ``only if''
  part of the proposition follows trivially from this.

  Conversely, suppose $\derive{f}_1, \derive{f}_2$ are derivatives for $f$, fix
  $a \in A$ and $\change{a} \in \changes{A}$ and let
  \begin{gather*}
    b = f(a)\\
    (p, q) = \derive{f}_1(a, \change{a})\\
    (r, s) = \derive{f}_2(a, \change{a})
  \end{gather*}
  
  First, note that $\bot \twist (p, q) = p = p \twist (p, q)$ and hence $\bot \kernel_{(p, q)} p$,
  therefore $\bot \kernel_{(r, s)} p$ since $(p, q) \kernelOrder (r,s)$. This implies
  \begin{displaymath}
    r = \bot \twist (r, s) =  p \twist (r, s) = p \vee r \wedge \neg s
  \end{displaymath}
  i.e. $r = p \vee r \wedge \neg s$ and hence $p \leq r \vee s$. Since $r$ and $s$ are disjoint,
  in order to prove $p \leq r$ it suffices to show that $p \wedge s = \bot$. 

  To see that this is the case, we note that 
  \begin{align*}
    p \wedge s &\leq b \twist (p, q)\\
               &=b \twist (r, s)\\
               &\leq \neg s
  \end{align*}
  Hence $p \wedge s \leq \neg s$, therefore $p \wedge s = \bot$, which concludes
  the proof that $p \leq r$. The proof that $q \leq s$ is symmetric.
\end{proof}

\subsection {Derivatives of Datalog formulae}
\concreteDatalog*
\begin{proof}
  \label{prf:concreteDatalog}
  Let $T$ be a Datalog formula with free relation variables $R_1, \ldots, R_n$,
  a choice of a semantics for the free relation variables $\semR_1, \ldots, \semR_n \in \Rel_\Gamma$
  and differences $\change{\semR_1}, \ldots, \change{\semR_n} \in \Rel_\Gamma \disjointTimes \Rel_\Gamma$.

  For brevity, we refer to the tuple $(\semR_1, \ldots, \semR_n)$ as $\semR$ and
  the tuple $(\change{\semR_1}, \ldots, \change{\semR_n})$ as $\diffR$. We abuse the notation
  and refer to $(\semR_1 \twist \change{\semR_1}, \ldots, \semR_n \twist \change{\semR_n})$
  as $\semR \twist \diffR$. We will also omit the arguments to $\upsem{T}$ and
  $\downsem{T}$ often, as there is never room for ambiguity.

  Then, it is the case that 
  \begin{displaymath}
    \sem{T}_\Gamma(\semR \twist \diffR)
    =
    \sem{T}_\Gamma(\semR) \twist 
    (\sem{\updiff{T}}_\Gamma(\diffR), 
    \sem{\downdiff{T}}_\Gamma(\diffR))
  \end{displaymath}

  We proceed by structural induction on $T$. We omit the cases for $\top$, $\bot$
  and relational variables for being trivial.
  
  Conjunction:
  \begin{align*}
    &\sem{T \wedge U}_\Gamma(\semR \twist \diffR)\\
    &= \sem{T}_\Gamma(\semR \twist \diffR) \cap \sem{U}_\Gamma(\semR \twist \diffR) \tag{semantics of $\wedge$}\\
    &= \sem{T}_\Gamma(\semR) \twist (\upsem{T}_\Gamma, \downsem{T}_\Gamma) 
      \cap \sem{U}_\Gamma(\semR) \twist (\upsem{U}_\Gamma, \downsem{U}_\Gamma) \tag{induction hypothesis}\\
    &= (\sem{T}_\Gamma(\semR) \cup \upsem{T}_\Gamma) \cap \neg \downsem{T}_\Gamma 
      \cap (\sem{U}_\Gamma(\semR) \cup \upsem{U}_\Gamma) \cap \neg \downsem{U}_\Gamma \tag{definition of $\twist$}\\
    &= (\sem{T}_\Gamma(\semR) \cap \neg \downsem{T}_\Gamma \cup \upsem{T}_\Gamma)
      \cap (\sem{U}_\Gamma(\semR) \cap \neg \downsem{U}_\Gamma \cup \upsem{U}_\Gamma) \tag{disjointness of $\upsem{}, \downsem{}$}\\
    &= (\sem{T}_\Gamma(\semR) \cap \neg \downsem{T}_\Gamma) \cap (\sem{U}_\Gamma(\semR) \cap \neg \downsem{U}_\Gamma) \tag{Boolean algebra}\\
    & \cup ((\upsem{T}_\Gamma \cap (\sem{U}_\Gamma(\semR) \cup \upsem{U}_\Gamma) \cap \neg \downsem{U}_\Gamma)
      \cap (\upsem{U}_\Gamma \cap (\sem{T}_\Gamma(\semR) \cup \upsem{T}_\Gamma) \cap \neg \downsem{T}_\Gamma))\\
    &= (\sem{T}_\Gamma(\semR) \cap \neg \downsem{T}_\Gamma) \cap (\sem{U}_\Gamma(\semR) \cap \neg \downsem{U}_\Gamma) \tag{induction hypothesis on $\bothdiff$}\\
    & \cup ((\upsem{T}_\Gamma \cap \bothsem{U}_\Gamma(\semR, \diffR))
            \cup (\upsem{U}_\Gamma \cap \bothsem{T}_\Gamma(\semR, \diffR)))\\
    &= \sem{T}_\Gamma(\semR) \cap \sem{U}_\Gamma(\semR) \cap \neg (\downsem{T}_\Gamma \cup \downsem{U}_\Gamma) \tag{Boolean algebra}\\
      &\cup ((\upsem{T}_\Gamma \cap \bothsem{U}_\Gamma(\semR, \diffR))
            \cup (\upsem{U}_\Gamma \cap \bothsem{T}_\Gamma(\semR, \diffR)))\\
    &= \sem{T}_\Gamma(\semR) \cap \sem{U}_\Gamma(\semR) \tag{Boolean algebra}\\
    & \cup ((\upsem{T}_\Gamma \cap \bothsem{U}_\Gamma)
            \cup (\upsem{U}_\Gamma \cap \bothsem{T}_\Gamma))
      \cap \neg (\downsem{T}_\Gamma \cup \downsem{U}_\Gamma)\\
    &= \sem{T \wedge U}_\Gamma(\semR)
      \twist (\upsem{(T \wedge U)}_\Gamma, \downsem{T}_\Gamma \cup \downsem{U}_\Gamma) \tag{definition of $\twist, \updiff$}\\
    &= \sem{T \wedge U}_\Gamma(\semR)
      \twist (\upsem{(T \wedge U)}_\Gamma, 
      (\downsem{T}_\Gamma \cap \sem{U}_\Gamma(\semR)) \cup (\downsem{U}_\Gamma \cap \sem{T}_\Gamma(\semR))) \tag{Boolean algebra}\\
    &= \sem{T \wedge U}_\Gamma(\semR)
      \twist (\upsem{(T \wedge U)}_\Gamma, \downsem{(T \wedge U)}_\Gamma) \tag{definition of $\downdiff{}$}
  \end{align*}
  
  Disjunction: identical to conjunction.
  
  Negation:
  \begin{align*}
    &\sem{\neg T}_\Gamma(\semR \twist \diffR)\\
    &= \neg \sem{T}_\Gamma(\semR \twist \diffR) \tag{semantics of $\neg$}\\
    &= \neg ((\sem{T}_\Gamma(\semR) \cup \upsem{T}_\Gamma) \cap \neg \downsem{T}_\Gamma) \tag{induction hypothesis, definition of $\twist$}\\
    &= \neg (\sem{T}_\Gamma(\semR) \cup \upsem{T}_\Gamma) \cup \downsem{T}_\Gamma \tag{Boolean algebra}\\
    &= \neg \sem{T}_\Gamma(\semR) \cap \neg \upsem{T}_\Gamma \cup \downsem{T}_\Gamma \tag{Boolean algebra}\\
    &= (\neg \sem{T}_\Gamma(\semR) \cup \downsem{T}_\Gamma) \cap \neg \upsem{T}_\Gamma \tag{disjointness of $\upsem{}, \downsem{}$}\\
    &= \sem{\neg T}_\Gamma(\semR) \twist (\downsem{T}_\Gamma, \upsem{T}_\Gamma) \tag{definition of $\twist$}\\
    &= \sem{\neg T}_\Gamma(\semR) \twist (\upsem{\neg T}_\Gamma, \downsem{\neg T}_\Gamma) \tag{definition of $\updiff{}, \downdiff{}$}
  \end{align*}

  Existential:
  Before proving the existential, we state the following (trivial) properties of the selection map $\sigma_\Gamma$:
  \begin{prop}[Distributivity of $\sigma_\Gamma$]
    \label{prop:sigmaDist}
    For any $A, B \in \Rel_{\Gamma, x}$ we have:
    \begin{align*}
      \sigma_\Gamma(A \cup B) &= \sigma_\Gamma(A) \cup \sigma_\Gamma(B)\\
      \sigma_\Gamma (A \cap \neg B) &= \sigma_\Gamma(A) \cap \neg \sigma_\Gamma(B) \cup \sigma_\Gamma(A \cap \neg B)
    \end{align*}
  \end{prop}
  
  With these in mind, we can proceed with the proof proper:
  \begin{align*}
    &\sem{\exists x . T}_\Gamma(\semR \twist \diffR)\\
    &=\sigma_\Gamma(\sem{T}_{\Gamma, x}(\semR \twist \diffR)) \tag{semantics of $\exists$}\\
    &=\sigma_\Gamma(\sem{T}_{\Gamma, x}(\semR) \twist (\upsem{T}_{\Gamma, x}, \downsem{T}_{\Gamma, x})) \tag{induction hypothesis}\\
    &=\sigma_\Gamma((\sem{T}_{\Gamma, x}(\semR) \cup \upsem{T}_{\Gamma, x}) \cap \neg \downsem{T}_{\Gamma, x}) \tag{definition of $\twist$}\\
    &=\sigma_\Gamma(\sem{T}_{\Gamma, x}(\semR) \cup \upsem{T}_{\Gamma, x}) \tag{\cref{prop:sigmaDist}}\\
    &  \cap \neg \sigma_\Gamma(\downsem{T}_{\Gamma, x})
      \cup \sigma_\Gamma((\sem{T}_{\Gamma, x}(\semR) \cup \upsem{T}_{\Gamma, x}) \cap \downsem{T}_{\Gamma, x})\\
    &=(\sigma_\Gamma(\sem{T}_{\Gamma, x}(\semR)) \cup \sigma_\Gamma(\upsem{T}_{\Gamma, x})) \tag{\cref{prop:sigmaDist}}\\
    &\cap \neg \sigma_\Gamma(\downsem{T}_{\Gamma, x})
      \cup \sigma_\Gamma((\sem{T}_{\Gamma, x}(\semR) \cup \upsem{T}_{\Gamma, x}) \cap \downsem{T}_{\Gamma, x})
      \\
    &=(\sigma_\Gamma(\sem{T}_{\Gamma, x}(\semR)) \cup \sigma_\Gamma(\upsem{T}_{\Gamma, x})) \tag{Boolean algebra}\\
    & \cap \neg 
      (\sigma_\Gamma(\downsem{T}_{\Gamma, x} \cap \neg \sigma_\Gamma((\sem{T}_{\Gamma, x}(\semR) \cup \upsem{T}_{\Gamma, x}) \cap \downsem{T}_{\Gamma, x})))
    \\
    &=(\sem{\exists x . T}_{\Gamma}(\semR) \cup \upsem{\exists x . T}_{\Gamma})
      \cap \neg 
      (\downsem{\exists x . T}_{\Gamma} \cap \neg \bothsem{\exists x . T}_\Gamma) \tag{definition of $\updiff{}, \downdiff{}$}\\
    &= \sem{\exists x . T}_\Gamma \twist (\upsem{\exists x . T}_\Gamma, \downsem{\exists x . T}_\Gamma) \tag{definition of $\twist$}
  \end{align*}

\end{proof}

\subsection{Directed-complete partial orders and fixpoints}

\booleanAlgebraContinuous*
\begin{proof}
  \label{prf:booleanAlgebraContinuous}
  $L$ is a complete lattice, so certainly a dcpo. $\cstr{L}_\superpose$ is a
  dcpo with $\bigvee (p_i, q_i) \defeq (\bigvee p_i \wedge \neg \bigwedge q_i, \bigwedge q_i)$.

  Continuity of $\twist$ in its second argument:
  \begin{align*}
    &a \twist \bigvee (p_i, q_i)\\
    &= a \twist (\bigvee p_i, \bigwedge q_i)\\
    &= (a \vee (\bigvee p_i \wedge \neg \bigwedge q_i)) \wedge \neg \bigwedge q_i\\
    &= (a \vee \bigvee p_i) \wedge \neg \bigwedge q_i\\
    &= (a \vee \bigvee p_i) \wedge \bigvee \neg q_i \tag{applying de Morgan}\\
    &= \bigvee (a \vee p_i) \wedge \neg q_i \tag{$\vee$ and $\wedge$ are continuous}\\
    &= \bigvee a \twist (p_i, q_i)
  \end{align*}

  Continuity $\twist$ in its first argument and continuity of $\splus$ follow easily from their definitions and the continuity
  of $\vee$ and $\wedge$.
\end{proof}

\factoringFixpoints*
\begin{proof}
  \label{prf:factoringFixpoints}
  Let
  \begin{displaymath}
    p(b) = (\lfp(f), g(\lfp(f), b))
  \end{displaymath}
  Then $h(h^i(\bot)) \leq p(p^i(\bot))$ (by simple induction), and so by continuity
  \begin{displaymath}
    \lfp(h) = \bigsqcup_{i \in \NN} h^i(\bot) \leq \bigsqcup_{i \in \NN} p^i(\bot) = \lfp(p)
  \end{displaymath}

  But $h(\lfp(p)) = \lfp(p)$, so $\lfp(h) \leq \lfp(p)$, since $\lfp(h)$ is least.

  Hence $\lfp(h) = \lfp(p) = (\lfp(f), \lfp(\lambda b . g(\lfp(f), b)))$.
\end{proof}

\iterDerivativesN*
\begin{proof}
  \label{prf:iterDerivativesN}
  By induction on $n$. We show the inductive step.
  \begin{align*}
    &\partial_2\iter(f, (n+1) + m)\\
    &=f(\iter(f, n + m)) \tag{definition of $\iter$}\\
    &=f(\iter(f, n) \cplus \partial_2{\iter}(f, n, m)) \tag{by induction}\\
    &=\iter(f, n+1) \cplus \derive{f}(\iter(f, n), \partial_2{\iter}(f, n, m)) \tag{$f$ is differentiable, definition of $\iter$}
  \end{align*}
\end{proof}

\fixpointIter*
\begin{proof}
  \label{prf:fixpointIter}
  \begin{align*}
    &\lfp(\iter_f)\\
    &=\bigsqcup_{n \in \NN} \iter(f, n)\\
    &=\bigsqcup_{n \in \NN} \pi_1 (\nextiter_f^n(\bot))
  \end{align*}
\end{proof}

\fixpointPseudoDerivatives*
\begin{proof}
  \label{prf:fixpointPseudoDerivatives}
  
  We show that a change $\change{w} \in \changes{A}$ satisfies
  the equation:
  \begin{equation}\label{eqn:fixcondition}
    \change{w} = \adjust(f, \change{f})(\change{w})
  \end{equation}
  if and only if $\fixpoint(f) \cplus \change{w}$ is a fixpoint of $f \cplus \change{f}$.

  Let $\change{w} \in \changes{A}$ satisfy \cref{eqn:fixcondition}. Then
  \begin{align*}
    &(f \cplus \change{f})(\fixpoint_A(f) \cplus \change{w})\\
    &= f(\fixpoint(f))
    \cplus
    \adjust(f, \change{f})(\change{w})
    \tag{by \cref{def:functionalChanges}}\\
    &= \fixpoint(f)
    \cplus
    \change{w}
    \tag{rolling the fixpoint and \cref{eqn:fixcondition}}
  \end{align*}

  Hence $\fixpoint(f) \cplus \change{w}$ is a fixpoint of $f \cplus \change{f}$. The converse
  follows from reversing the direction of the proof.
\end{proof}

\iterDerivativesF*
\begin{proof}
  \label{prf:iterDerivativesF}
  The base case is easy to prove.

  For the inductive step:
  \begin{align*}
    &\iter(f \cplus \change{f}, n + 1)\\
    &=(f \cplus \change{f})(\iter(f \cplus \change{f}, n))\\
    &= (f \cplus \change{f})(
        \iter(f, n)
        \cplus \partial_1{\iter}(f, \change{f}, n)
      )
    \tag{ by induction}\\
    &= f(\iter(f, n)) \cplus \derive{\ev}((f, \iter(f, n)), (\change{f},
      \partial_1{\iter}(f, \change{f}, n)))
    \tag{by \cref{def:functionalChanges}}\\
    & =\iter(f, n + 1) \cplus \partial_1{\iter}(f, \change{f}, n)
  \end{align*}
\end{proof}

\leastFixpointDerivatives*
\begin{proof}
  \label{prf:leastFixpointDerivatives}
  $\partial_1{\iter}$ and $\nextiter_{f, \change{f}}$ are continuous since
  $\derive{\ev}$ and $f$ are.

  Hence the set $\{ \partial_1{\iter}(\cdot, \cdot, n) \}$ is directed, and so $\bigsqcup_{i \in \NN}
  \{ \partial_1{\iter}(\cdot, \cdot, n) \}$ is indeed a derivative for $\lfp$.

  We now show that it is equivalent to $\derive{\lfp}$:
  \begin{align*}
    &\bigsqcup_{n \in \NN} \partial_1{\iter}(f, \change{f}, n)\\
    &=\bigsqcup_{n \in \NN} \pi_2(\nextiter_{f, \change{f}}^n(\bot))\\
    &=\pi_2(\bigsqcup_{n \in \NN} \nextiter_{f, \change{f}}^n(\bot)) \tag{$\pi_2$ is continuous}\\
    &= \pi_2 (\lfp(\nextiter_{f, \change{f}})) \tag{$\nextiter_{f, \change{f}}$ is continuous, Kleene's Theorem}\\
    &= \pi_2 ((\lfp(f), \lfp (\lambda\ \change{a}. \derive{\ev}((f, \lfp f), (\change{f}, \change{a})))))
    \tag{by \cref{prop:factoringFixpoints}, and the definition of $\nextiter$}\\
    &= \pi_2 (\lfp(f), \lfp(\adjust(f, \change{f})))\\
    &= \lfp(\adjust(f, \change{f}))\\
    &= \derive{\lfp}(f, \change{f})
  \end{align*}
\end{proof}

\end{document}